\documentclass[fleqn,usenatbib]{mnras}
\usepackage[T1]{fontenc}

\DeclareRobustCommand{\VAN}[3]{#2}
\let\VANthebibliography\thebibliography
\def\thebibliography{\DeclareRobustCommand{\VAN}[3]{##3}\VANthebibliography}

\usepackage{graphicx}	
\usepackage{amsmath}	
\usepackage{amssymb}	
\usepackage[utf8]{inputenc}
\usepackage{color}
\usepackage{tabularx}
\usepackage{newtxtext,newtxmath}

\newcommand{\ud}{\mathrm{d}}

\title[MHD Code Comparison]{A Comparison of 2D Magnetohydrodynamic Supernova Simulations with the \textsc{CoCoNuT-FMT} and \textsc{Aenus-Alcar} Codes}

\author[Varma et al.]{
Vishnu Varma,$^{1}$\thanks{E-mail:vishnu.rvejayan@monash.edu}
Bernhard M\"uller$^{1}$\thanks{E-mail:bernhard.mueller@monash.edu}
and Martin Obergaulinger$^{2}$
\\
$^{1}$School of Physics and Astronomy, 19 Rainforest Walk, Monash University, VIC 3800, Australia\\
$^{2}$Departament d'Astonomia i Astrof\'isca, Universitat de Val\`encia, Edifici d'Investigatci\'o Jeroni Munyoz, C/Dr.\ Moliner, 50,\\
\phantom{$^8$}E-46100 Burjassot (Val\`encia),
Spain
}
\date{Accepted XXX. Received YYY; in original form ZZZ}

\pubyear{2021}

\begin{document}
\label{firstpage}
\pagerange{\pageref{firstpage}--\pageref{lastpage}}
\maketitle

\begin{abstract}
Code comparisons are a valuable tool for the verification of supernova simulation codes and the quantification of model uncertainties. Here we present a first comparison of axisymmetric magnetohydrodynamic (MHD) supernova simulations with the \textsc{CoCoNuT-FMT} and \textsc{Aenus-Alcar} codes, which use distinct methods for treating the MHD induction equation and the neutrino transport. We run two sets of simulations of a rapidly rotating $35M_\odot$ gamma-ray burst progenitor model with different choices for the initial field strength,  namely $10^{12}\,\mathrm{G}$ for the maximum poloidal and toroidal field in the strong-field case and $10^{10}\,\mathrm{G}$ in the weak-field case. We also investigate the influence of the Riemann solver and the resolution in \textsc{CoCoNuT-FMT}. The dynamics is qualitatively similar for both codes and robust with respect to these numerical details, with a rapid magnetorotational explosion in the strong-field case and a delayed neutrino-driven explosion in the weak-field case. Despite relatively similar shock trajectories, we find sizeable differences in many other global metrics of the dynamics, like the explosion energy and the magnetic energy of the proto-neutron star. Further differences emerge upon closer inspection, for example, the disk-like surface structure of the proto-neutron star proves highly sensitives to numerical details. The electron fraction distribution in the ejecta as a crucial determinant for the nucleosynthesis is qualitatively robust, but the extent of neutron- or proton-rich tails is sensitive to numerical details. Due to the complexity of the dynamics, the ultimate cause of model differences can rarely be uniquely identified, but our comparison helps gauge uncertainties inherent in current MHD supernova simulations.
\end{abstract}

\begin{keywords}
supernovae: general-- MHD -- methods: numerical
\end{keywords}



\section{Introduction}\label{intro}
It was first recognised by \citet{Baade1934a} that in supernova explosions of massive stars ($\mathord{\geq} 8 \mathrm{M_\odot}$) a small fraction of the energy released during the terminal collapse
of the stellar iron core is somehow transferred to the stellar envelope to produce a powerful explosion known as a core-collapse supernova (CCSN). The ensuing quest for the details of the explosion mechanism has been ongoing for decades. 
Among the many suggested explosion scenarios, the neutrino-driven mechanism, initially proposed by \citet{Colgate1966}
has emerged as the favoured explanation for the majority of CCSN explosions. In the modern paradigm of the delayed
 neutrino-driven mechanism \citep{Bethe1985}, shock revival is accomplished thanks to the increase of the post-shock pressure due to partial reabsorption of neutrinos that stream out from the young proto-neutron star (PNS) and the cooling layer of accreted material on its surface.
 Detailed simulations have determined that additional support by multi-dimensional instabilities like turbulent convection \citep{Herant1995,Burrows1995} and the standing accretion  shock instability \citep{Blondin2003} is critical to make the neutrino-driven mechanism work.

Partly due to increasing computing power, supernova simulations have been able to continually include more detailed physics such as multi-group neutrino transport, magnetic fields, and  general relativity, and to abandon symmetry assumptions in favour of fully three-dimensional models, but even the most sophisticated
simulations still need to make  approximations
(for recent reviews, see
 \citealt{Janka2012_review,Burrows2013,Muller_2016review,Mueller_LRCA,Mezzacappa_LRCA}).
The complexity of the codes and the disparities in numerical methodology present a challenge for drawing robust conclusions on the workings of the supernova engine from the models. 
Code comparison studies are a useful approach for increasing the robustness and reliability of simulation results and have already addressed many important ingredients of supernova codes, some focusing more strictly on the neutrino transport \citep[e.g.][]{Yamada1999, Ott2008, Richers2017, Chan2020}, whereas others have more broadly addressed the interplay of neutrino transport and hydrodynamics 
in 1D, 2D \citep{Liebendorfer2005, Mueller2010, OConnor2018, Just2018} and to some extent in  3D \citep{Cabezon2018} with different solvers for the hydrodynamics and the neutrino transport alike. Numerical uncertainties related to the effects of grid resolution have also received considerable attention \citep[e.g.,][]{Abdikamalov2014,Handy2013,Hanke2012,Couch2014a,Radice2015,Nagakura2019,Melson2020a}.

A similarly broad exploration of numerical uncertainties is yet to be performed for  magnetorotational explosions, which constitute  the best-explored alternative to the neutrino-driven mechanism and are widely thought to
operate in hyperenergetic supernovae (``hypernovae''), some of which are also associated with long gamma ray bursts (GRBs; \citealp{Woosley2006b}).
In this scenario, whose history can be traced to early pioneering work of \citet{Leblanc1970,Bisnovatyi-Kogan1976,Meier1976,Mueller1979}, rapid rotation and strong magnetic fields are the critical ingredients for the explosion; the magnetic fields help tap
the rotational energy of the forming compact object  and channel it into kinetic energy in outflows -- usually in the form of collimated, magnetically-dominated jets -- which are driven either by magnetic forces and pressure or through viscous heating behind the shock.

With the inclusion  of rotation and magnetic fields in models, many additional unknowns enter into the simulations and have received substantial attention. The magnetorotational explosion mechanism is intimately linked to the problem of rotation and magnetism in massive stars. Initial conditions for magnetorotational explosion simulations currently come from ``1.5D'' stellar evolution models that assume shellular rotation and include effective recipes for magnetic field generation and angular momentum transport by hydrodynamic and magnetohydrodynamic processes \citep{Heger2000,Heger2005,Woosley2006}. Efforts to better understand the angular momentum distribution and magnetic fields in the cores of massive stars by means of multi-dimensional simulation have only started recently \citep{Varma2021,Yoshida2021,McNeill2021}. Since 1.5D stellar evolution models predict magnetic fields that are rather weak and predominantly toroidal, the general notion has long been that field amplification processes after collapse are critical in magnetorotational explosions, although this has recently been challenged \citep{Obergaulinger2017,Obergaulinger2021,Aloy2021}. These amplification processes have duly received considerable attention.
\citet{Akiyama2003} pointed out the potential importance of the magnetorotational instability \citep[MRI;][]{Balbus1991}, which has been addressed in the context of CCSNe in a host of numerical and analytic studies
\citep{Obergaulinger2009,Sawai2013,Sawai2015,Masada2015,Guilet2015a,Guilet2015b,Moesta2015,Rembiasz2016,Rembiasz2016b,Reboul2021}. Field amplification by an $\alpha$-$\Omega$ dynamo in the PNS \citep{Thompson1993} has also been investigated in multi-D simulations as an avenue for generating strong magnetic fields on large scales from weak seed fields
\citep{Raynaud2020}.
While important issues like the initial fields and angular momentum distribution, the saturation of the MRI, and non-ideal effects still warrant further investigation, global simulations of magnetorotational explosions based on parameterised initial fields
have been available for quite some time in 2D both with \citep{Burrows2007, Dessart2007,Obergaulinger2017,Obergaulinger2020,Bugli2020,Aloy2021,Jardine2021} and without
\citep{Kotake2004,Takiwaki2004,Sawai2005,Obergaulinger2006,Moiseenko2006,Takiwaki2009} detailed neutrino transport, and are becoming routine in 3D \citep{Winteler2012,Moesta_2014,Moesta_2017,Halevi2018}, even though only few models \citep{Kuroda2020,Obergaulinger2021,Bugli2021} also include neutrino transport.

As magnetohydrodynamic supernova simulations mature and are being used to understand hypernova light curves \citep{Shankar2021} and explosion properties and the potential role of hypernovae as a site for the rapid-neutron capture process \citep{Winteler2012,Moesta_2017,Reichert2021}, it is critical to better address  numerical uncertainties in these models following a similar approach as for neutrino-driven explosions, where comparison studies with different codes and resolutions studies have done much to distinguish between robust and uncertain features of the mechanism. However, to date, no study has as  yet compared simulations of magnetorotational supernovae using independent codes.

In this paper, we present the first comparison of axisymmetric (2D) magnetorotational explosions simulations between two independent codes, namely the \textsc{CoCoNuT-FMT} \citep{Mueller2015,MullerVarma2020} and \textsc{Aenus-Alcar} \citep{Obergaulinger, Obergaulinger2014} codes.
We also conduct a minimal resolution study with \textsc{CoCoNuT-FMT} models and explore the sensitivity of
the results to the approximate Riemann solver. Our paper is organised as follows:  In Section~\ref{numMethods}, we describe the  magnetohydrodynamic (MHD) solvers in \textsc{CoCoNuT-FMT}
and \textsc{Aenus-Alcar} and also provide a brief description of the neutrino transport schemes. We then outline the simulation setup in Section~\ref{models}. We compare the \textsc{CoCoNuT-FMT}
and \textsc{Aenus-Alcar} results in Section~\ref{results} and present results for our minimal resolution study with \textsc{CoCoNuT-FMT}
in Section~\ref{ResAnalysis}. We conclude with a summary and evaluation in Section~\ref{conclusion}.

\section{Numerical Methods} \label{numMethods}

We perform axisymmetric core-collapse supernova simulations with the pseudo-Newtonian MHD version of \textsc{CoCoNuT-FMT} \citep{Mueller2010, Mueller2015} and the special relativistic \textsc{Aenus-Alcar} code \citep{Obergaulinger,Just2015}.
In both codes, we use an effective gravitational potential (Case~A from \citealt{Marek2006}) to approximately capture general relativistic effects. The nuclear equation of state of
\citet{Lattimer1991} with an incompressibility of
$K=220 \, \mathrm{MeV}$ is used for all models at
high densities. Here
we briefly outline the magnetohydrodynamic and neutrino transport solvers in the two codes. For more details, we refer to the relevant code papers.

\subsection{\textsc{CoCoNuT-FMT} Code}
The \textsc{CoCoNuT-FMT} supernova code is built on a Godunov-based finite-volume solver for
spherical polar coordinates that uses higher-order reconstruction \citep{Colella1984,Colella2008}. It has recently been upgraded to solve the Newtonian MHD equations \citep{MullerVarma2020} using the HLLC Riemann solver \citep{Gurski2004,Miyoshi2005}, and the divergence free condition $\nabla\cdot\mathbf{B} = 0$ is maintained using a variant of hyperbolic divergence cleaning introduced in \cite{Dedner2002}. This method involves introducing an additional scalar field, $\psi$, known as a generalized Lagrange multiplier, which couples to the induction equation and propagates the divergence errors at a cleaning speed $c_\mathrm{h}$ to the domain boundaries.  We apply a modification of hyperbolic divergence cleaning that allows for variable cleaning speeds following the ideas of \cite{Tricco2016} for increased stability. 
Compared to the original cleaning method, we instead propagate a modified scalar field,
$\hat{\psi}=\psi/c_\mathrm{h}$, where the cleaning speed $c_\mathrm{h}$ is chosen as one fourth of the maximum local signal speed allowed by the Courant-Friedrichs-Lewy limit 
(which is different from \citealt{MullerVarma2020},
where the cleaning speed was set to the fast magnetosonic wave speed).
 The extended system of MHD equations for the density $\rho$, velocity $\mathbf{v}$, magnetic field $\mathbf{B}$, the total energy density $e$, and the Lagrange multiplier $\hat{\psi}$ reads,
\begin{eqnarray}
\partial_t \rho
+\nabla \cdot \rho \mathbf{v}
&=&
0,
\\
\partial_t (\rho \mathbf v)
+\nabla \cdot \left(\rho \mathbf{v}\mathbf{v}-
\frac{\mathbf{B} \mathbf{B}}{4\pi}
+P_\mathrm{t}\mathcal{I}
\right)
&=&
\rho \mathbf{g}
\\
\nonumber
&&
-
\frac{(\nabla \cdot\mathbf{B}) \mathbf{B}}{4\pi}
+
 \mathbf{Q_m}
,
\\
\partial_t e+
\nabla \cdot 
\left[(e+P_\mathrm{t})\mathbf{u}
-\frac{\mathbf{B} (\mathbf{v}\cdot\mathbf{B})}{4\pi}
\right]
&=&
\rho \mathbf{g}\cdot \mathbf{v}
\\
\nonumber
&&
+
Q_e + \mathbf{Q_m}\cdot\mathbf{v}
-\mathbf{B}\cdot \nabla (c_\mathrm{h}\psi)
,
\\
\partial_t \mathbf{B} +\nabla \cdot (\mathbf{v}\mathbf{B}-\mathbf{B}\mathbf{v})
+\nabla  (c_\mathrm{h} \hat{\psi})
&=&0
\\
\partial_t \hat{\psi}
+c_\mathrm{h} \nabla \cdot \mathbf{B}
&=&-\hat{\psi}/\tau.
\end{eqnarray}
where $\mathbf{g}$ is the gravitational acceleration, $P_\mathrm{t}$ is the total pressure, $\mathcal{I}$ is the Kronecker tensor, $c_\mathrm{h}$ is the hyperbolic cleaning speed, $\tau$ is the damping timescale for divergence cleaning. The neutrino energy and momentum source terms are denoted by  $Q_\mathrm{e}$ and $\mathbf{Q}_\mathrm{m}$. 
Different from \citet{MullerVarma2020}, the total energy density  $e$ does not contain the cleaning field
and is given by its standard definition,
\begin{equation}
e=\rho \left(\epsilon+\frac{v^2}{2}\right)+\frac{B^2}{8\pi},
\end{equation}
where $\epsilon$ is the mass-specific internal energy. 

In the \text{CoCoNuT} models the innermost $10\, \mathrm{km}$ are treated in 1D to avoid an
overly stringent Courant time step limit
near the origin of the grid. The thermodynamic
variables and the radial velocity are assumed
to be spherically symmetric. The non-radial
velocity components and the magnetic
fields are not set to zero in the core, however.
Instead, we assume uniform rotation at a
rate corresponding to the total angular
momentum in the core, and the magnetic
field in the core is represented by
vector spherical harmonics of degree $\ell=1$,
whose coefficients are evolved according
to the induction equation.

The fast multi-group transport (FMT) scheme
of \citet{Mueller2015} is used for the neutrino transport. The FMT scheme solves the energy-dependent stationary zeroth-momentum equation using a hybrid closure based on a
two-stream Boltzmann solution and an analytic Eddington closure at flux factors $h>0.5$. The scheme accounts for gravitational redshift, but mostly disregards velocity-dependent terms, except for an effective frame correction of the absorption opacity at low optical depth \citep{Mueller_2019}. The set of neutrino rates includes emission by and absorption on nucleons and nuclei, neutrino scattering on nucleons and nuclei, and, for heavy-flavour neutrinos, Bremsstrahlung in an effective one-particle approximation. Recoil energy transfer in neutrino-nucleon scattering is included approximately.
Nucleon potentials are taken into account in charged-current neutrino-nucleon interactions \citep{Martinez-Pinedo2012}. 
Nucleon correlations are included following
\citet{Horowitz2017}. Neutral-current scattering
cross sections are computed following
\citet{Horowitz2002} with a somewhat
large value for the nucleon
strangeness of $g_\mathrm{A}=-0.1$.
The deleptonisation scheme of \citet{Liebendoerfer2005b} is used during the collapse phase.

\subsection{\textsc{Aenus-Alcar} Code}
The \textsc{Aenus-Alcar} code solves the special relativistic version of the MHD equations.  Different from \textsc{CoCoNuT}, the divergence constraint is  accounted for by using a constrained transport scheme \citep{Evans1998}.
The equations of MHD are also solved using the HLLC solver, but combined
with MP5 reconstruction \citep{Suresh1997}.
The code can accommodate different coordinate systems, but is also run in spherical polar coordinates in this study.
 \textsc{Aenus-Alcar} treats the innermost $10\,\mathrm{km}$ in 2D, but with a coarser grid resolution.

Neutrinos are treated by a spectral scheme for two-moment transport with an analytic closure \citep[see][for an overview]{Just2015}. Neutrinos of different energies are coupled via the effects  of Doppler shifts, aberration, and gravitational redshift treated to $\mathcal{O}(v/c)$.
The set of reactions included in the \textsc{Aenus-Alcar} simulations is given in detail in, e.g., \citet{Obergaulinger2017,Obergaulinger2018} and consists of  emission/absorption and scattering reactions with nucleons and nuclei,  inelastic scattering off electrons, electron-positron annihilation, and nucleon-nucleon Bremsstrahlung.

We do not aim to reproduce perfectly identical
neutrino microphysics in both codes. We rather pit the two codes against each other as used in normal production runs so that the comparison is more reflective of ``real-life'' uncertainties encountered in magnetorotational supernovae. To give a conservative assessment of possible errors, we view it as one useful strategy to overestimate rather than to underestimate code differences by not making the models more similar than they would be in independent studies. Of course, this also entails some risk of accidental error cancellation in some (but not all) metrics of the code comparison. This  strategy is complementary to the opposite approach of reproducing all of the physics as closely as possible in different codes, which will be a useful undertaking for the future, perhaps within a comparison broader with a broader community base.

\section{Model Setup} \label{models}

All our simulations use the stellar progenitor model 35OC of \citet{Woosley2006}. This model was computed using
the \textsc{Kepler} code including a prescription for angular momentum transport by magnetic torques \citep{Spruit2002,Heger2005}. The 35OC stellar model is a rapidly rotating model with an initial total angular momentum of $J_\mathrm{init} = 10^{52}\,\mathrm{erg}\,\mathrm{s}$ and a surface rotational velocity $v_\mathrm{rot} = 380\, \mathrm{km/s}$ on  the zero-age main sequence (ZAMS). The progenitor has a ZAMS mass of $M_\mathrm{ZAMS} = 35 M_\odot$, but at the time of core collapse mass loss has reduced its total mass to $M_\mathrm{final} = 28.07 M_\odot$.

We set up our models using the shellular rotation profile of the progenitor, but
modify the original magnetic fields of the 35OC model to similar initial conditions as in models 35OC-Rs (strong field) and 35OC-Rw (weak field) from \citet{Obergaulinger2017}. Strong field models
are set up with an initial maximum poloidal and toroidal magnetic field of $B_\mathrm{pol,tor} = 10^{12}\,\mathrm{G}$ and weak field cases have $B_\mathrm{pol,tor} = 10^{10}\,\mathrm{G}$. Throughout the rest of this paper, references to the strong-field models will include a suffix ``-Rs''  while the weak-field models will have a suffix ``-Rw'' to the model names.
The magnetic field geometry in either case is defined using the vector potential $\mathbf{A}$ from \citet{Suwa2007},
\begin{align}
\label{eq:magini}
    \left(A^{r}, A^{\theta}, A^{\phi}\right)=
\frac{1}{2\left(r^{3}+r_{0}^{3}\right)}
    \left(B_\mathrm{tor} r_{0}^{3} r \cos \theta, 0, B_\mathrm{pol} r_{0}^{3} r \sin \theta\right),
\end{align}
where the radius parameter is set to $r_0=10^8\,\mathrm{cm}$.\footnote{Note that models 35OC-Rs and 35OC-Rw in \citet{Obergaulinger2017} had inadvertently used a slightly different radial dependence for the toroidal field, which is why \textsc{Alcar} models were rerun
with the exact initial conditions given
in Equation~(\ref{eq:magini}).
}

All the \textsc{CoCoNuT-FMT} models 
(denoted by a prefix ``coco'') have a computational domain that extends out to $R_\mathrm{out} = 10^{5}\, \mathrm{km}$ 
using a nonuniform radial grid of 550 zones
and spans colatitudes $\theta$ in the range $[0^{\circ}, 180^{\circ}]$ on a spherical polar grid with 128 zones in $\theta$. The \textsc{Aenus-Alcar} models 
(denoted by a prefix ``Alcar'')
implement the same angular resolution, but have a computational domain of 400 radial zones that extends to $R_\mathrm{out} =1.4 \times 10^{5}\, \mathrm{km}$.

We simulate three pairs of models with \textsc{CoCoNuT-FMT}. 
For each strong- and weak-field model, we conduct runs using  the MHD version of the HLLC Riemann solver \citep{Gurski2004} or the HLLE solver \citep{Einfeldt1988}. These runs are labelled as coco-HLLC-Rs/w and coco-HLLE-Rs/w, respectively.
In addition, we conduct two high-resolution runs
coco-Res-Rs/w with the HLLC solver with 256 zones in angle
for a minimal resolution test. The parameters of all runs are summarised in Table~\ref{tab:models}.

\begin{table*}
	\centering
    \resizebox{\textwidth}{!}{
    \begin{tabular}{c c c c c c c}
     \hline
     \textbf{Model} & \textbf{Riemann solver}  & \textbf{Resolution}  & \textbf{Maximum initial field $B_\mathrm{pol}, B_\mathrm{tor}$} &
     \textbf{Time of bounce} & \textbf{Final simulation time} &
     \textbf{Explosion Energy}\\ [0.5ex] 
     \hline
     Alcar-Rs & HLLC & $400 \times 128$ zones & $10^{12}\, \mathrm{G}$ & 0.433\,s & 0.397\,s & $1.27\times 10^{51}\,\mathrm{erg}$ \\
     Alcar-Rw & HLLC & $400 \times 128$ zones& $10^{10}\, \mathrm{G}$ & 0.430\,s & 0.870\,s & $1.11\times 10^{50}\,\mathrm{erg}$\\
     coco-HLLC-Rs & HLLC & $550 \times 128$ zones& $10^{12}\, \mathrm{G}$ & 0.396\,s & 0.516\,s & $7.71\times 10^{50}\,\mathrm{erg}$\\
     coco-HLLC-Rw & HLLC & $550 \times 128$ zones & $10^{10}\, \mathrm{G}$ & 0.396\,s & 1.005\,s & $2.42\times 10^{50}\,\mathrm{erg}$\\
     coco-HLLE-Rs & HLLE & $550 \times 128$ zones& $10^{12}\, \mathrm{G}$ & 0.395\,s & 0.482\,s & $6.69\times 10^{50}\,\mathrm{erg}$\\
     coco-HLLE-Rw & HLLE & $550 \times 128$ zones& $10^{10}\, \mathrm{G}$ & 0.395\,s & 0.948\,s & $5.19\times 10^{50}\,\mathrm{erg}$\\
     coco-Res-Rs & HLLC & $550 \times 256$ zones& $10^{12}\, \mathrm{G}$ & 0.396\,s & 0.434\,s & $5.49\times 10^{50}\,\mathrm{erg}$\\
     coco-Res-Rw & HLLC & $550 \times 256$ zones& $10^{10}\, \mathrm{G}$ & 0.396\,s & 0.924\,s & $2.09\times 10^{50}\,\mathrm{erg}$\\
     \hline
    \end{tabular} }
    \caption{Overview of the eight models included in the comparison, summarising the setup (code, Riemann solver, resolution, initial magnetic field strength) as well as the time of bounce, the final post-bounce simulation time, and the explosion energy for each run, recorded at $\mathord{\approx} 0.85\, \mathrm{s}$ for the weak-field models (``-Rw'' ) and
    at  $\mathord{\approx} 0.39\, \mathrm{s}$ for the strong-field models (``-Rs'').
    Riemann solvers and resolution of all models run on CoCoNuT and the ones run with \textsc{Aenus-Alcar}.}
    \label{tab:models}
\end{table*}

\section{Results -- Code Comparison} \label{results}

In the following, we present a comparison of
the standard resolution models coco-HLLC-Rs, coco-HLLC-Rw, coco-HLLE-Rs and coco-HLLE-Rw to models Alcar-Rs and Alcar-Rw as described in Table \ref{tab:models}. We discuss the differences between the \textsc{Aenus-Alcar} and \textsc{CoCoNuT-FMT} results, as well as between the HLLC and HLLE Riemann solvers implemented in \textsc{CoCoNuT-FMT}. 

\subsection{Shock Propagation} \label{shock_propagation}
\begin{figure}
   \centering
  \includegraphics[width=\linewidth]{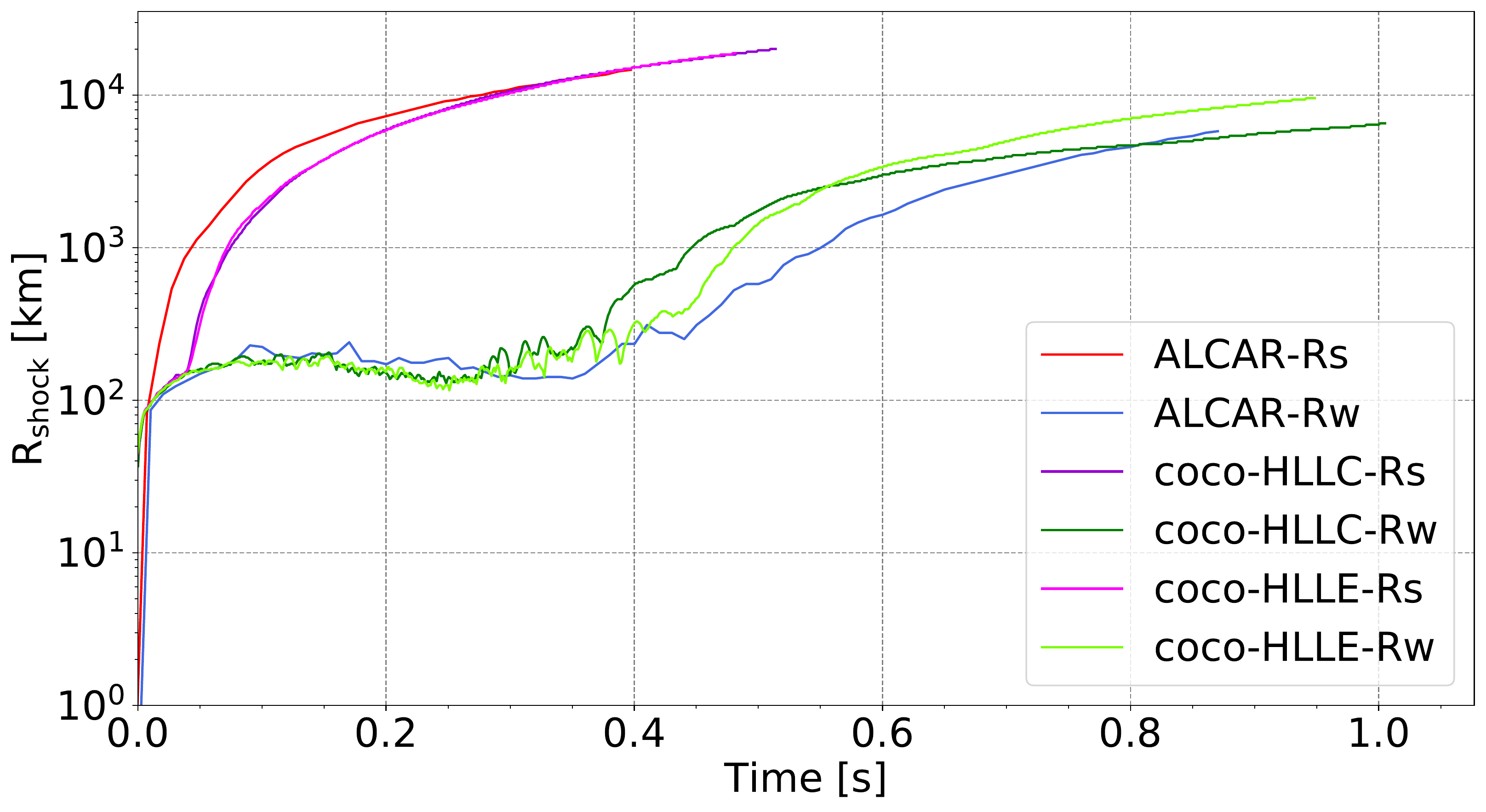}
  \caption{Evolution of the maximum
  shock radius at standard
  resolution for \textsc{CoCoNuT-FMT} using the HLLC and HLLE Riemann solver models along with the \textsc{Aenus-Alcar} models. Both the 
  strong-field (-Rs) and weak-field (-Rw)
  models  are shown.
  }
  \label{fig:ShockRadius}
\end{figure}

We begin by comparing the propagation of
the shock (Figure~\ref{fig:ShockRadius})
between the different models as this is
one of the most basic metrics for the
post-bounce dynamics. Due to the strong asymmetry of the explosion,
 the dynamics is best captured by
the maximum shock radius rather than the angle-averaged
shock radius, especially in the case of the strong magnetic field models, which develop a veritable jet-driven explosion with strongly collimated polar outflows.

In Alcar-Rs, shock expansion starts promptly after  core bounce and does not exhibit a stalled shock phase. By contrast, both coco-HLLC-Rs and coco-HLLE-Rs stall for a short period of $\mathord{\approx} 50\,\mathrm{ms}$ before the shock is revived. After shock revival, the shock in
 models  coco-HLLC-Rs and coco-HLLE-Rs expands rapidly, achieving similar maximum shock radii  as Alcar-Rs at the end of the simulations.

\begin{figure}
    \centering
	\includegraphics[width=\linewidth]{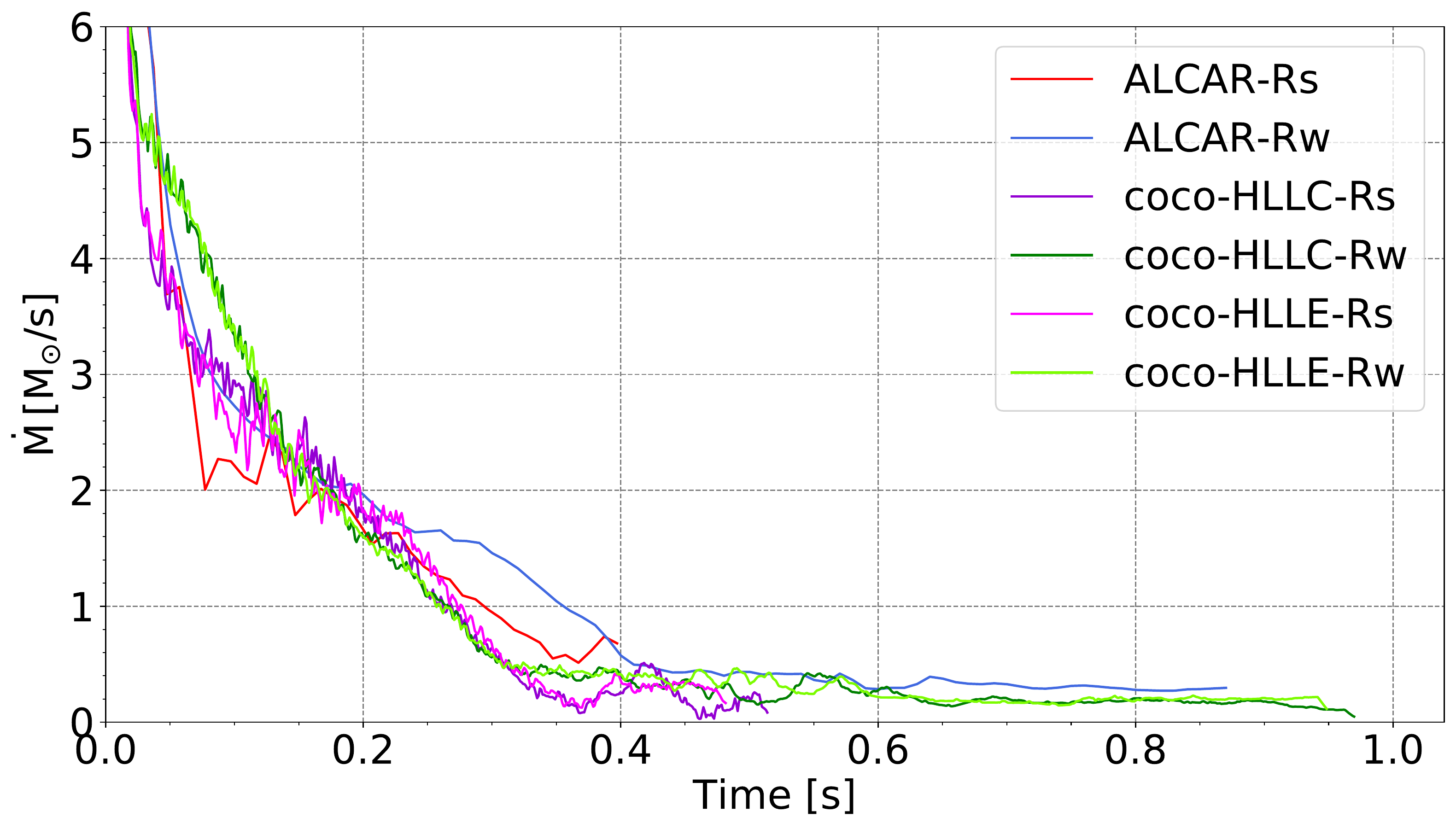}
\caption{Mass accretion rate
  $\dot{M}$  as a function of post-bounce time, measured as the
  time derivative of the mass of the PNS (defined  as the region where 
  $\rho > 10^{11} \,\mathrm{g}\,\mathrm{cm}^{-3}$).
  }
\label{fig:accretion}
\end{figure}

The different time of explosion in \textsc{Alcar} and \textsc{CoCoNuT}  may be due to different dynamics of magnetic field amplification in the PNS surface region that result in the earlier emergence of a jet in model Alcar-Rs. There may, however, be a compounding factor in the form of the collapse dynamics. Both in the strong-field case and the weak-field case, the Alcar models take $30\texttt{-}35\, \mathrm{ms}$ longer to collapse. The mass accretion rate at $45\texttt{-}130\, \mathrm{ms}$ differs noticeably between
\textsc{CoCoNuT} and \textsc{Alcar} with
\textsc{CoCoNuT} exhibiting higher mass
accretion rates early on after collapse (Figure~\ref{fig:accretion}).
The collapse time is known to be sensitive
to the microphysics \citep{huedepohl_phd,Powell2021}, and although
both codes use the same high-density equation
of state, they differ in their treatment
of the low-density equation of state, nuclear burning, and deleptonisation during collapse. Factors that affect the collapse dynamics may be exacerbated by transient adjustment of the stellar models after mapping the 1D stellar evolution model with shellular rotation into the supernova simulation codes. Although the resulting difference of the infall dynamics are mostly confined to the collapse phase, it is known that differences in the mass accretion rate can persist for some time after collapse even in simulations of non-rotating progenitors \citep{huedepohl_phd}. In the subsequent analysis, one must bear in mind that this will inevitably impact the shock propagation as well as the evolution of the PNS properties, and, more importantly, the  neutrino emission at early post-bounce times to some degree.

Unlike the strong-field models, the models with weaker magnetic fields (coco-HLLC-Rw, coco-HLLE-Rw, Alcar-Rw) all go through a prolonged accretion phase, similar to
the dynamics in purely neutrino-driven explosions
of massive progenitors. The shock trajectories during
the accretion phase are similar, with Alcar-Rw exhibiting  slightly larger stalled shock radii early on, but slightly slower shock expansion later. Most of the time, the
relative difference between the runs remains
below $10\%$.
All three models undergo shock revival around $0.4\, \mathrm{s}$ post-bounce with minor differences.
The Alcar-Rw model is the last
to explode and continues to have a smaller shock radius than both  \textsc{CoCoNuT-FMT} models until about $0.8\, \mathrm{s}$ where its shock radii slightly surpasses that of coco-HLLC-Rw.
At their largest divergence, the relative difference between the
three models is $\mathord{\approx} 12\,\%$.
We also note that the \textsc{CoCoNuT-FMT} weak-field models show a much more pronounced and prolonged 
activity of the standing accretion shock instability
(SASI; \citealp{Blondin2003, Foglizzo2007,Guilet2012},
visible as oscillations in the maximum shock radius in coco-HLLC-Rw and coco-HLLE-Rw
in the phase leading up to shock revival. 
This is in line with the established finding
that stronger shock retraction is conducive
to more vigorous SASI activity \citep{Scheck2008,Mueller2012b}.

Shock revival in model coco-HLLE-Rw occurs
about $\mathrm{0.05s}$ later than in model coco-HLLC-Rw, but the shock subsequently expands faster than in the HLLC model, exhibiting a larger maximum shock radius at the end of the simulation runs. By contrast, no significant differences are seen between coco-HLLE-Rs and coco-HLLC-Rs. 
Since shock trajectories differ between the HLLE
and HLLC run only in the weak-field case, which develops
a ``hybrid'' explosion supported alike by magnetic
fields, neutrino heating, and convection \citep{Obergaulinger2017}, 
we suspect the differences between  coco-HLLE-Rw  and coco-HLLC-Rw do not actually reflect a systematic
dependence on the Riemann solver, but are due to stochastic variations, which have been shown to be considerable
in explosion models without MHD  in 2D
\citep{Muller2015a,Cardall2015a}.

Overall, the differences in shock propagation between
the corresponding \textsc{CoCoNuT-FMT} and \textsc{Aenus-Alcar} models are within
the range of deviations between independent codes
in comparisons of neutrino hydrodynamics simulations
in 2D \citep{Mueller2015,Cabezon2018,Just2018}.
Even in controlled 1D comparisons
with identical neutrino rates
\citep{Liebendorfer2005,Mueller2010,OConnor2018},
deviations of the order of $10\texttt{-}20\%$
in shock radius are not uncommon.
We suspect that the prompt explosion in Alcar-Rs that is not mirrored in the strong magnetic field models of \textsc{CoCoNuT-FMT} is an artefact of the different treatments of the inner core. As described in Section~\ref{models}, \textsc{Aenus-Alcar} treats the inner core in coarse 2D geometry, which allows for rotational shear and hence more rapid magnetic field amplification immediately after bounce, and consequently, larger magnetic pressure to immediately drive the shock out. These difference
in magnetic field amplification will be discussed  further in Section~\ref{PNS}.

\subsection{Explosion Energetics \label{explosion_dynamics}}

\begin{figure}
  \includegraphics[width=\linewidth]{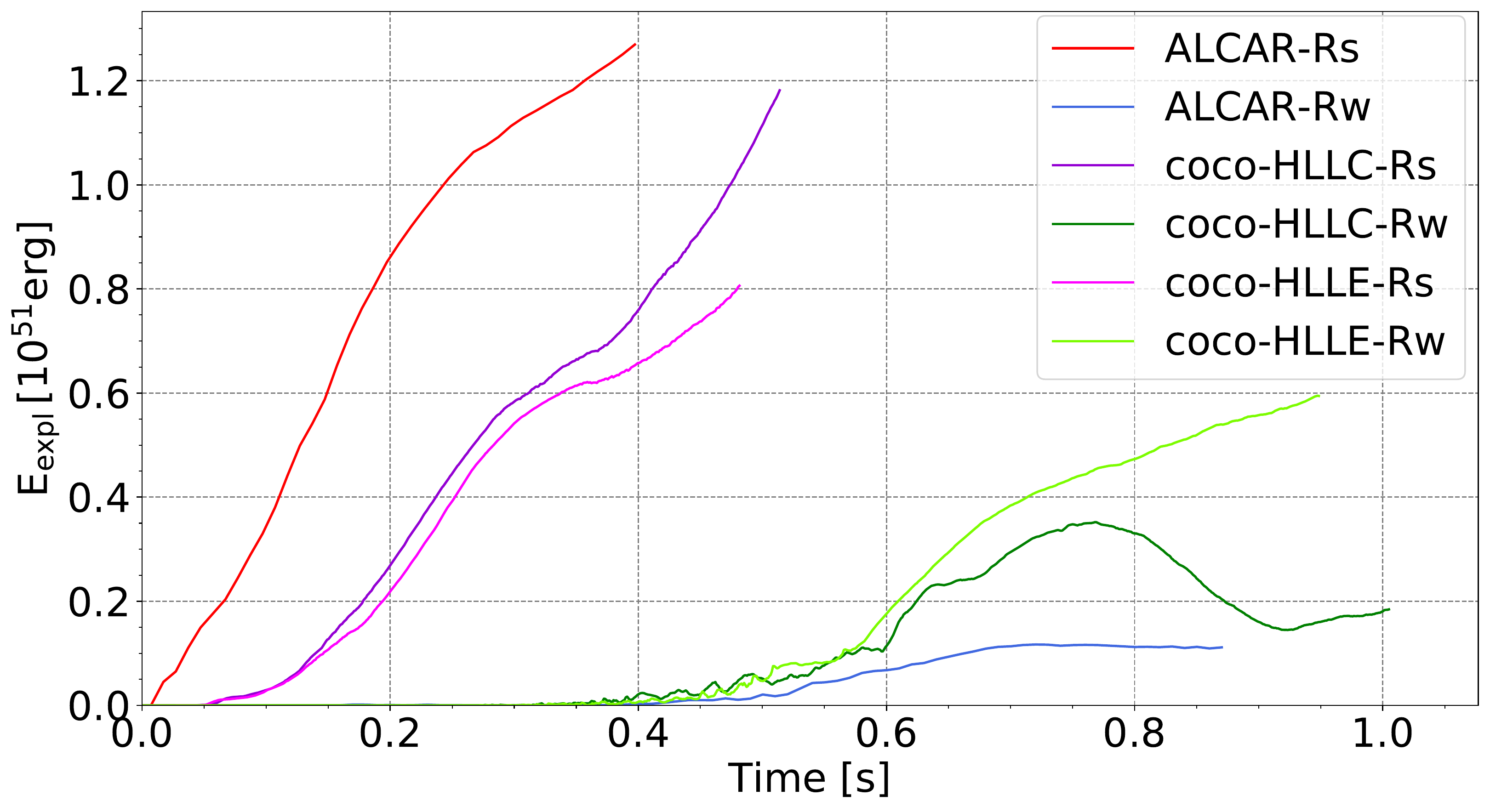}
  \caption{Evolution of the diagnostic explosion energy 
  $E_\mathrm{expl}$ for \textsc{CoCoNuT-FMT} and \textsc{Aenus-Alcar} models with strong and weak initial magnetic fields. }
  \label{fig:ExplEnergy}
\end{figure}

As the next metric of comparison for the global
dynamics, we consider the diagnostic explosion energy
$E_\mathrm{expl}$
\citep{Buras2006}. The diagnostic explosion energy 
is defined as an integral over the region that is nominally 
unbound,
\begin{equation}
E_\mathrm{expl} = \int\limits_{e_\mathrm{tot}>0} \rho e_\mathrm{tot}
\,\ud V ,
\end{equation}
where $e_\mathrm{tot}$ is the total energy density, i.e., the sum of the internal, kinetic, gravitational, and magnetic energy density (the latter of which was not included in \citealt{Buras2006}).
The evolution of the diagnostic explosion energy $E_\mathrm{expl}$
is shown for all models  in Figure~\ref{fig:ExplEnergy}.
The strong magnetic field models exhibit a very quick rise in explosion energy
following the early and rapid expansion of the shock, whereas the weak-field models develop positive $E_\mathrm{expl}$ at $ \mathord{\sim} 400 \, \mathrm{ms}$ after bounce, and remain below $10^{51}\, \mathrm{erg}$ in explosion energy.

Although the \textsc{CoCoNuT-FMT} and \textsc{Aenus-Alcar} models show qualitative
agreement in the evolution of $E_\mathrm{expl}$, 
Figure~\ref{fig:ExplEnergy} more clearly exhibits
quantitative differences between the runs than could
be seen in the shock propagation. Among the strong-field
models, models coco-HLLE-Rs and coco-HLLC-Rs agree
quite well with each  other, but develop a positive diagnostic
energy later than the Alcar-Rs run, which is consistent with
the longer delay to shock revival discussed in the previous 
section.
After this slight delay, model coco-HLLC-Rs in particular exhibits comparable rates of explosion energy growth, with Alcar-Rs increasing at $\mathord{\approx} 3.12 \times 10^{51}\, \mathrm{erg}\, \mathrm{s}^{-1}$ while model coco-HLLC-Rs shows,
on average, a rise at a rate of $\mathord{\approx} 3.17 \times 10^{51}\, \mathrm{erg}\,\mathrm{s}^{-1}$. 
The explosion in coco-HLLE-Rs 
initially grows at a similar rate as in coco-HLLC-Rs, but the model becomes slightly less energetic by the end of the simulation. The cause of this divergence is unclear but since our simulations end by $500 \, \mathrm{ms}$, it is possible that this is just a temporary deviation due to some element of stochasticity in the dynamics of the evolving explosion.
The diagnostic energy is evidently a much more sensitive
measure for the dynamics than the shock radius. The energetics of the narrowly
collimated jets that drive shock expansion is sensitive
to a number of details, e.g., how efficiently
energy is extracted from differential rotation, and
how much material is entrained by the collimated jets that
drive the explosion. It should also be noted that
relatively minor differences in the kinetic, magnetic,
and thermal energy of the ejecta can translate into
larger differences once the potential energy is subtracted.
For these reasons, the rather large discrepancy 
in explosion energy between the \textsc{Aenus-Alcar} and \textsc{CoCoNuT-FMT}
models is not surprising.  In Section~\ref{PNS},
we shall demonstrate that the two codes produce
somewhat different PNS structures
that can motivate (though not rigorously explain)
significant differences in the energetics of the
jet-driven explosions. We will also revisit
the sensitivity of the explosion energy in our resolution analysis in Section~\ref{ResAnalysis}.

The weak-field models appear to be qualitatively more disparate than their strong-field counterparts. 
All three models have significant delays  of $\mathord{\sim} 400 \, \mathrm{ms}$ to the explosion with a slightly longer delay phase for the Alcar-Rw model, mirroring the different time of shock revival (Figure~\ref{fig:ShockRadius}). 
Although the shock radii of all three models are comparable by the end of the simulations, we find pronounced quantitative differences in their explosion energies; the
explosion energy of model Alcar-Rw is $80\%$ lower than that of coco-HLLC-Rw, while
model coco-HLLE-Rw ends up with a significantly
higher explosion energy than the other two models by  the end of the simulations. This does not indicate
fundamental differences between the codes, but
is likely just a reflection of the strongly stochastic
explosion dynamics of 2D models in the
``hybrid'' regime where neutrino heating,
convection or SASI, and magnetic fields jointly drive the explosion. The unsteady,
and even non-monotonic growth of the explosion
energy in coco-HLLC-Rw and Alcar-Rw
clearly reveals this stochasticity, and
the variations between the different runs
are similar to the variations seen
in non-magnetic models for different
realisations of the same model setup
with the same code \citep{Muller2015a,Cardall2015a}.
This does not, of course, preclude that there are other
systematic differences in the dynamics between the
weak-field runs that are hidden behind the stochastic
variations.

\subsection{Proto-Neutron Star Parameters and Magnetic Field Amplification}
\label{PNS}
\begin{figure}
  \includegraphics[width=\linewidth]{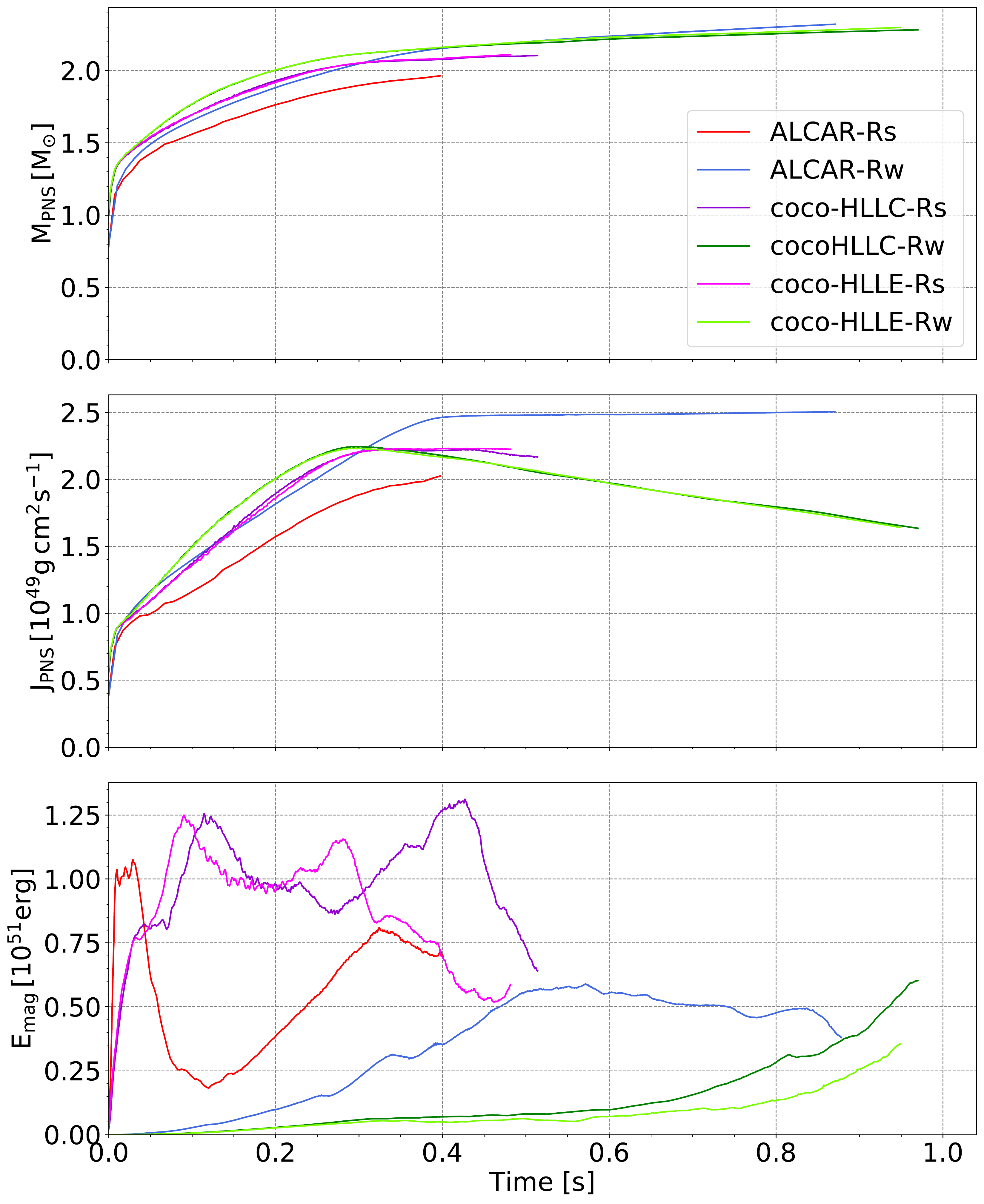}
  \caption{PNS mass
  $M_\mathrm{PNS}$, angular momentum
  $J_\mathrm{PNS}$, and magnetic energy 
  $E_\mathrm{mag}$,
  as a function of post-bounce time for the standard-resolution models. The PNS is defined as
  the region where $\rho > 10^{11} \,\mathrm{g}\,\mathrm{cm}^{-3}$.}
  \label{fig:PNS}
\end{figure}

\begin{figure}
    \includegraphics[width=\linewidth]{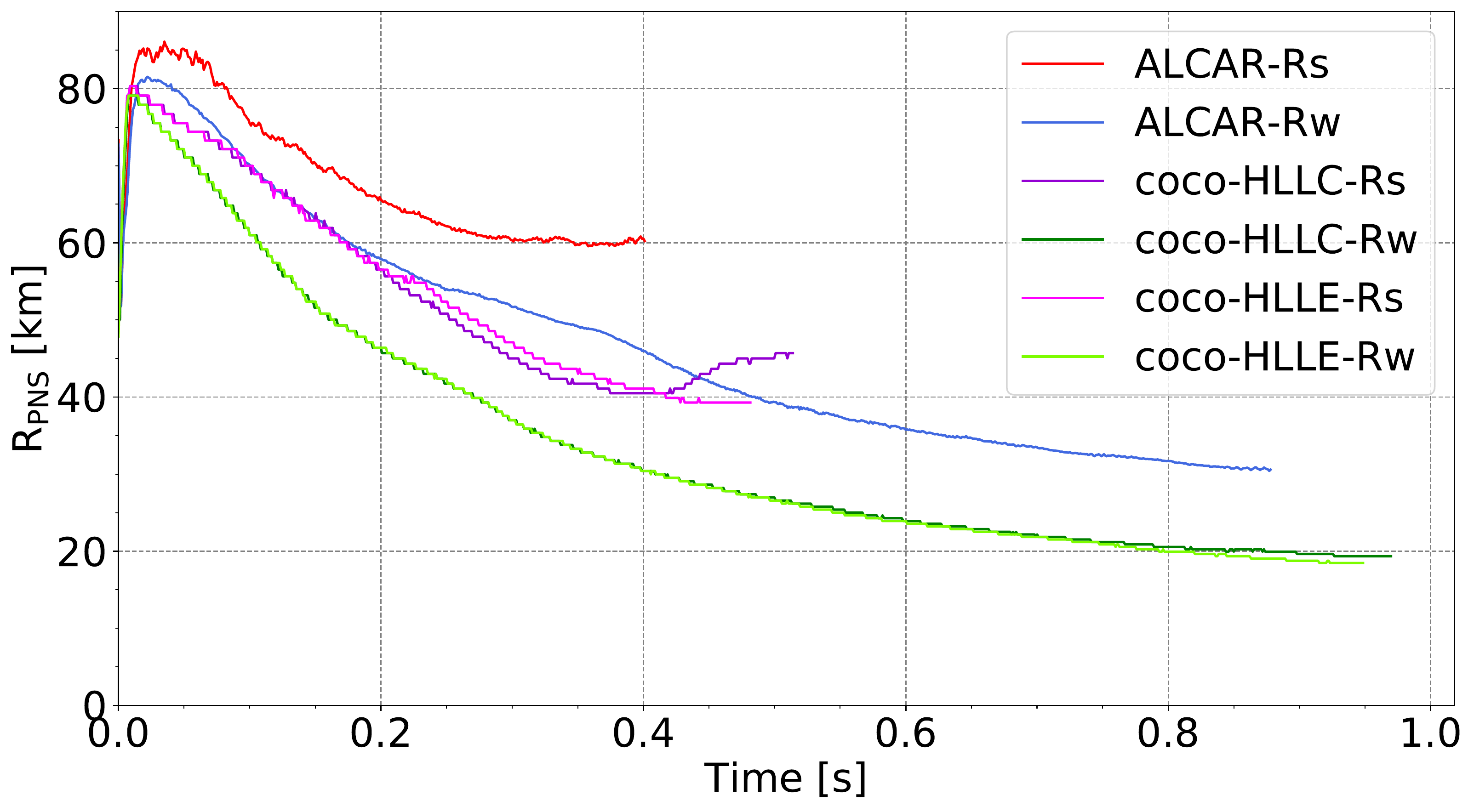}
  \caption{Comparison of the time evolution of the angle-averaged PNS radius
  $R_\mathrm{PNS}$ between the \textsc{CoCoNuT-FMT} and \textsc{Aenus-Alcar} models.}
  \label{fig:PNS_rad}
\end{figure}

We next consider the bulk parameters and structure of the PNS, and the
amplification of magnetic fields inside it
as key drivers of the overall dynamics.
Figure~\ref{fig:PNS} shows the
PNS mass $M_\mathrm{PNS}$, angular
momentum $J_\mathrm{PNS}$, and magnetic
energy $E_\mathrm{mag}$, which we
compute as volume integrals over
the region where
$\rho > 10^{11} \,\mathrm{g}\, \mathrm{cm}^{-3}$
following the customary, though somewhat
arbitrary definition of the PNS surface.
The evolution of the angle-averaged PNS radius
$R_\mathrm{PNS}$ for the different runs is plotted in Figure~\ref{fig:PNS_rad}.
Angle-averaged radii are determined
by identifying the PNS surface using the same fiducial density
$\rho=10^{11} \, \mathrm{g} \, \mathrm{cm}^{-3}$ based on angle-averaged density profiles.

The growth of the PNS mass (top panel in Figure~\ref{fig:PNS}) is very similar in the weak-field models Alcar-Rw, coco-HLLC-Rw and coco-HLLE-Rw runs. We do, however, see an extended period before $0.35s$ where the \textsc{CoCoNuT-FMT} models have a more massive PNS. 
The Alcar-Rw model then catches up after $0.35\,\mathrm{s}$, and eventually ends up with a slightly more massive PNS. 
Among the strong-field models, coco-HLLC-Rs and coco-HLLE-Rs exhibit more substantial accretion onto the PNS immediately after collapse compared to model Alcar-Rs, leading to a ~$9\%$ higher PNS mass in the \textsc{CoCoNuT-FMT} runs throughout. This is explained by the more rapid onset of the explosion seen in Alcar-Rs. The PNS mass difference largely accounts for a similar difference in PNS angular momentum between the 
\textsc{CoCoNuT-FMT} and
\textsc{Alcar} strong-field
models (Figure~\ref{fig:PNS}, middle panel).

More  dramatic differences between the two codes can be
seen in the total PNS angular momenta after about $0.3\,\mathrm{s}$ post-bounce for the weak-field models. Models coco-HLLC-Rw and coco-HLLE-Rw start to gradually lose angular momentum, whereas model Alcar-Rw continues to slowly gain angular momentum as it accretes mass. This difference is especially noteworthy since the PNS masses of the weak-field models are quite similar throughout. This behaviour in the \textsc{CoCoNuT-FMT} models is likely due to spurious non-conservation of angular momentum due to our numerical implementation in axisymmetry, 
where the velocity component
$v_\varphi$ is zeroed in the zones adjacent
to the grid axis. 
Aside from these boundaries, 
we shall see that the different treatments of the inner PNS 
lead to significant differences in the angular momentum profiles between the two codes. 

More pronounced differences between the \textsc{CoCoNuT-FMT} and  \textsc{Aenus-Alcar} models 
than in the PNS mass and angular momentum are found 
in the PNS magnetic energy
$E_\mathrm{mag}$
(bottom panel of Figure~\ref{fig:PNS}). In
coco-HLLC-Rs and coco-HLLE-Rs,
the magnetic energy in the PNS first peaks around
$1.25\times 10^{51}\, \mathrm{erg}$
at $0.1\, \mathrm{s}$ and remains high for some time,
decaying after a second peak. Although the overall time evolution
of $E_\mathrm{mag}$ in these two models is quite similar, coco-HLLE-Rs is set apart by an earlier gradual drop in magnetic energy around $0.1\mathrm{s}$ before  coco-HLLC-Rs. This is noteworthy, considering that the choice of Riemann solver shows very little effect on the other PNS quantities. In model Alcar-Rs, the
magnetic field energy peaks immediately after core bounce at $\mathord{\approx} 1.05\times 10^{51}\, \mathrm{erg}$, but then  decreases sharply to a much lower level. 

The weak magnetic field models exhibit similarly large differences as their strong-field counterparts. In particular, we see a much earlier rise in PNS magnetic energy in Alcar-Rw than in coco-HLLC-Rw and coco-HLLE-Rw, followed by a slow, gradual drop after $\mathord{\sim} 0.5\, \mathrm{s}$. By contrast, the weak-field runs with \textsc{CoCoNuT-FMT} show a gradual rise in magnetic energy over time that continues until the end of the simulation. 

\begin{figure}
  \includegraphics[width=\linewidth]{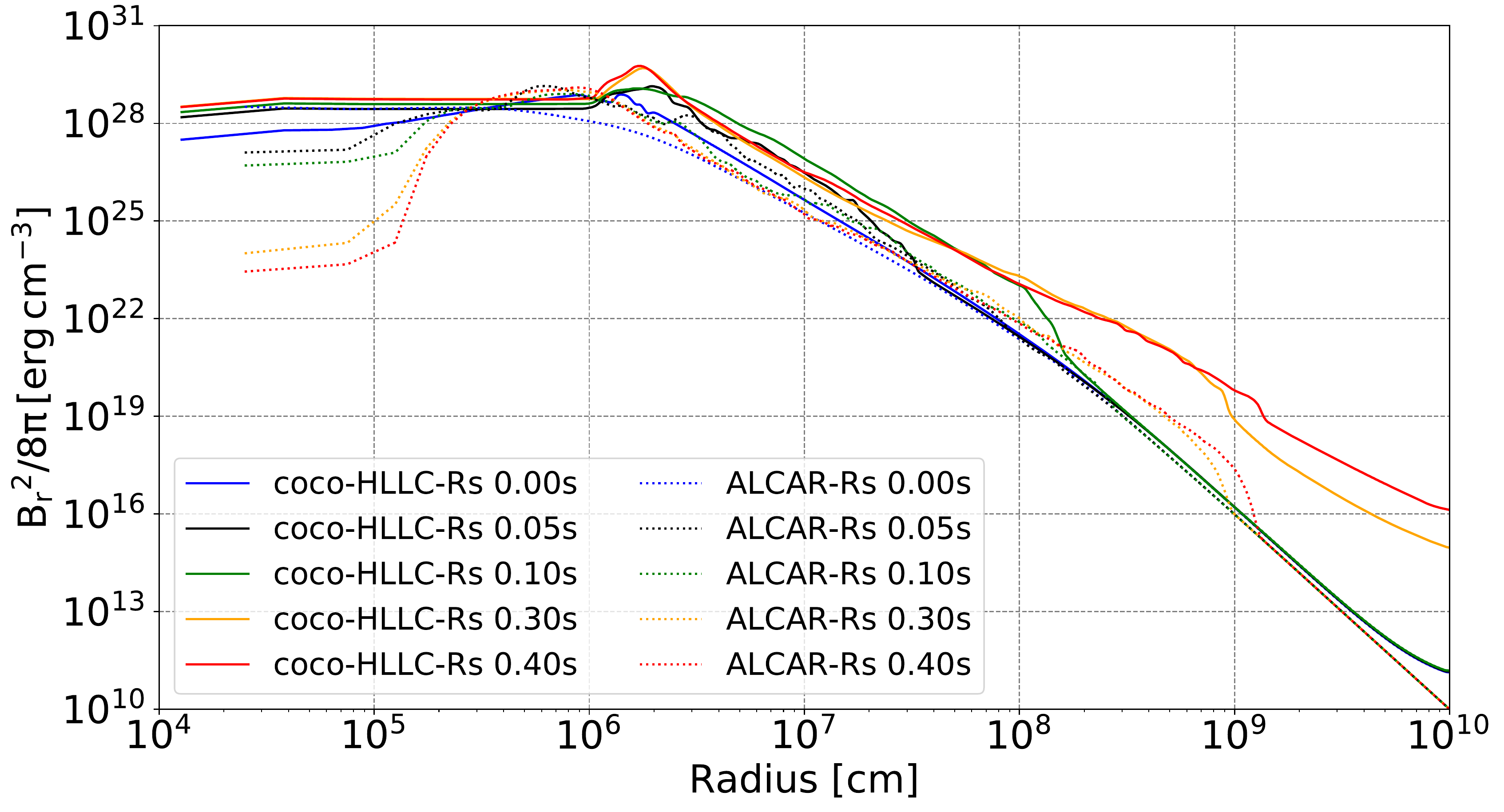}
   \includegraphics[width=\linewidth]{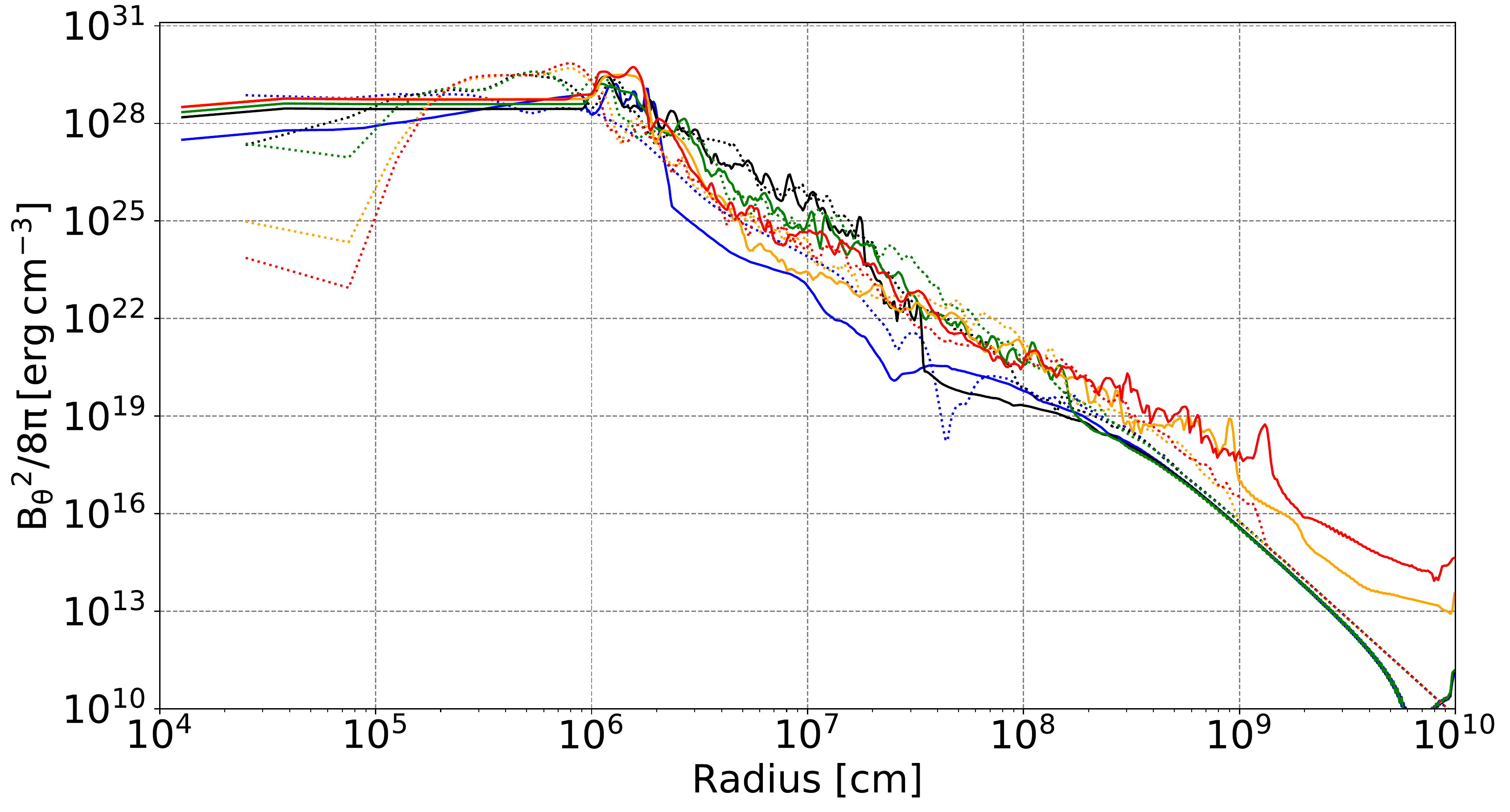}
   \includegraphics[width=\linewidth]{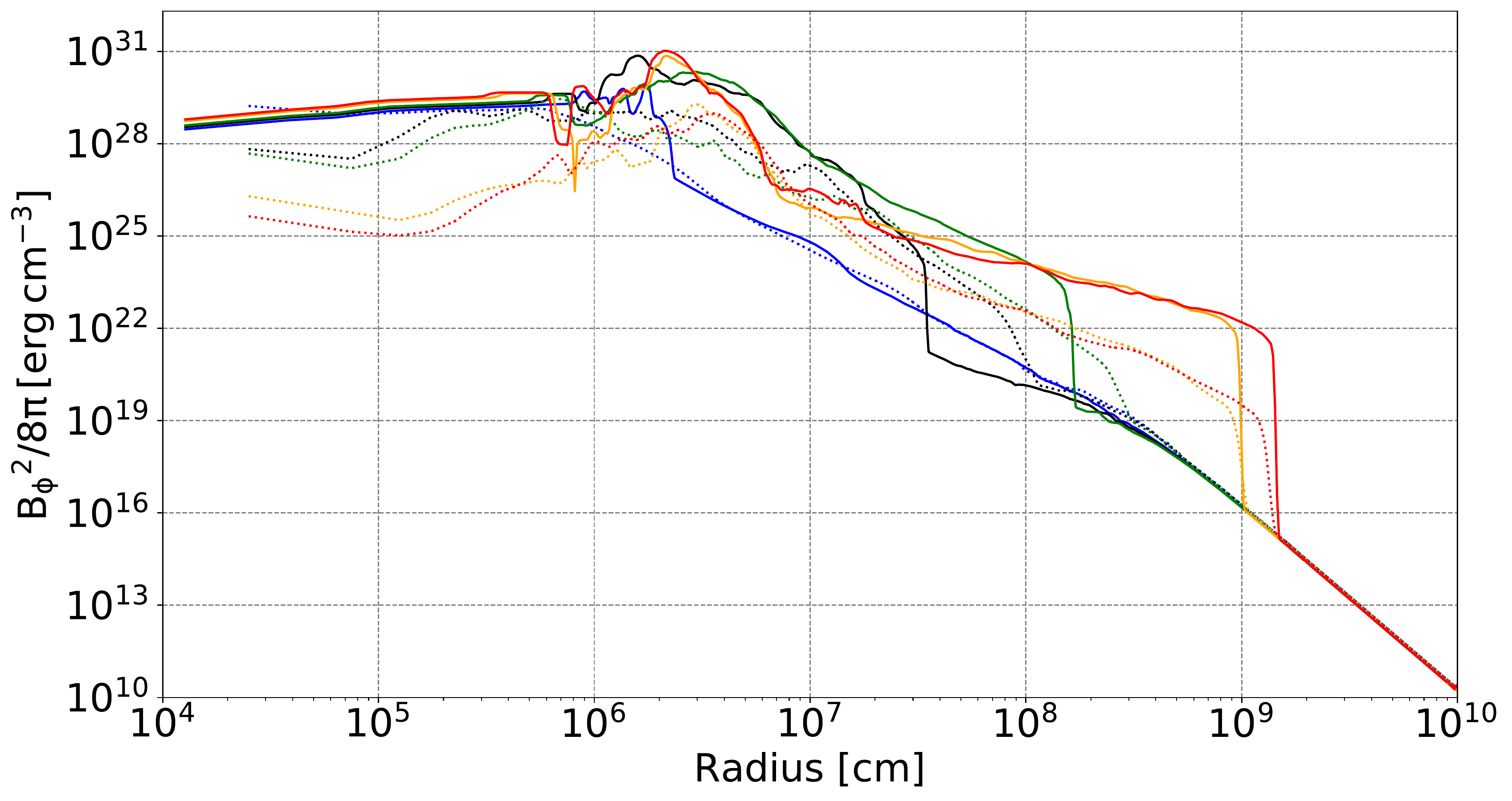}
 \caption{Top to bottom:
Profiles of the energy density
contained in the radial, meridional, and zonal components
$B_r$, $B_\theta$, and $B_\varphi$
of the magnetic field for the strong-field
models coco-HLLC-Rs (solid lines)
and  Alcar-Rs (dashed lines) 
at various post-bounce times.
}
\label{fig:MagProfile}
\end{figure}

\begin{figure}
  \includegraphics[width=\linewidth]{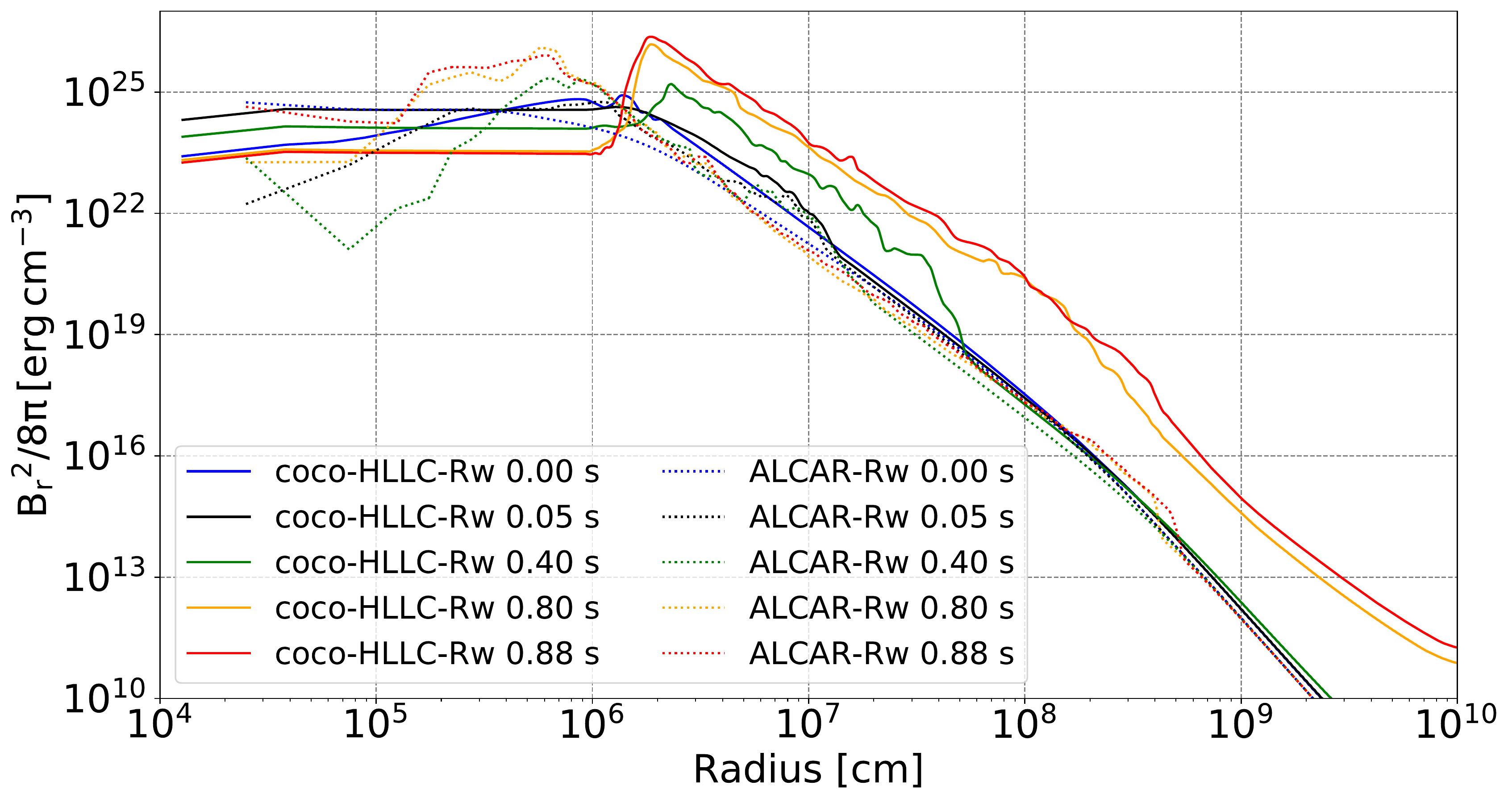}
  \includegraphics[width=\linewidth]{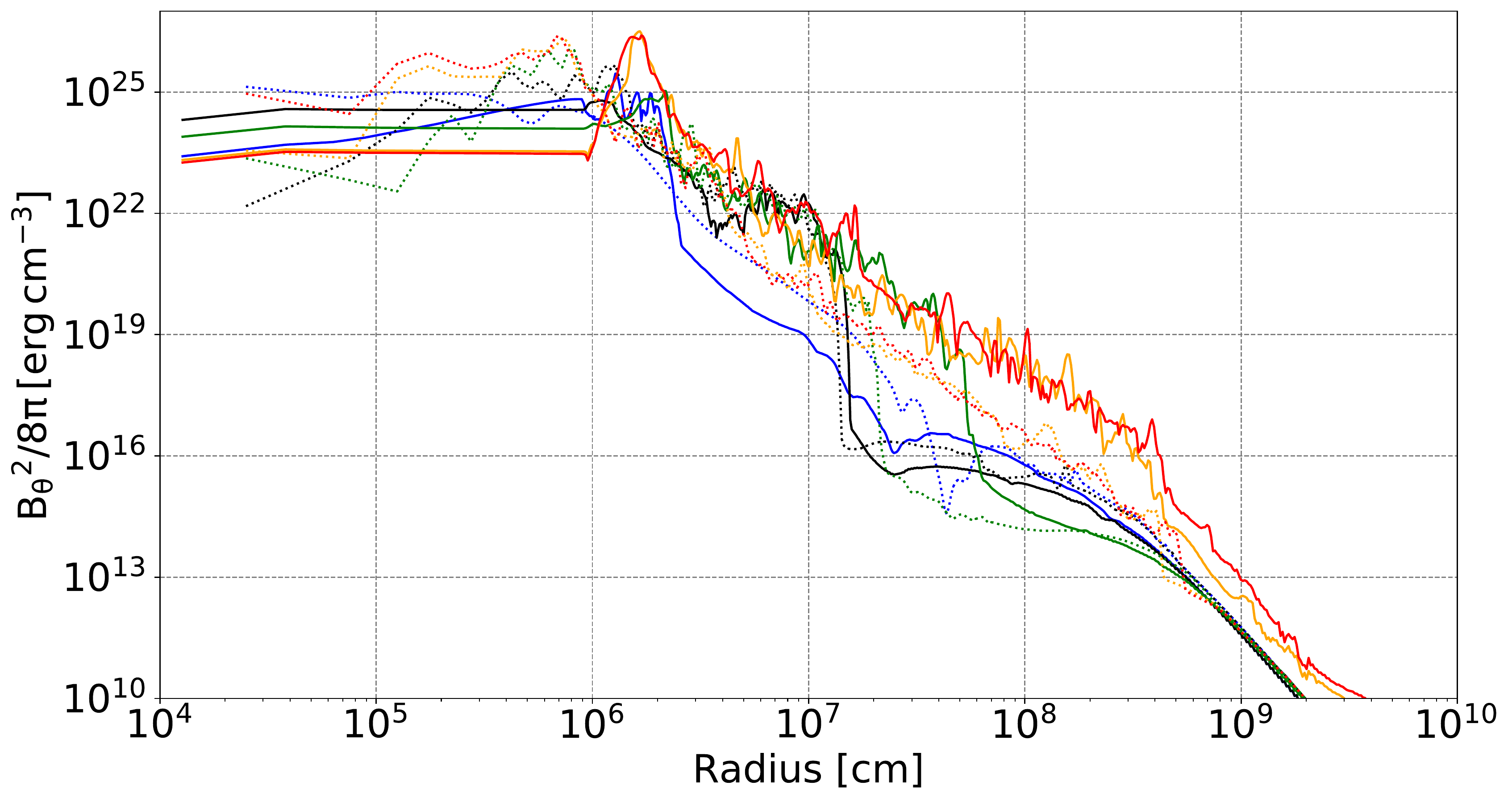}
  \includegraphics[width=\linewidth]{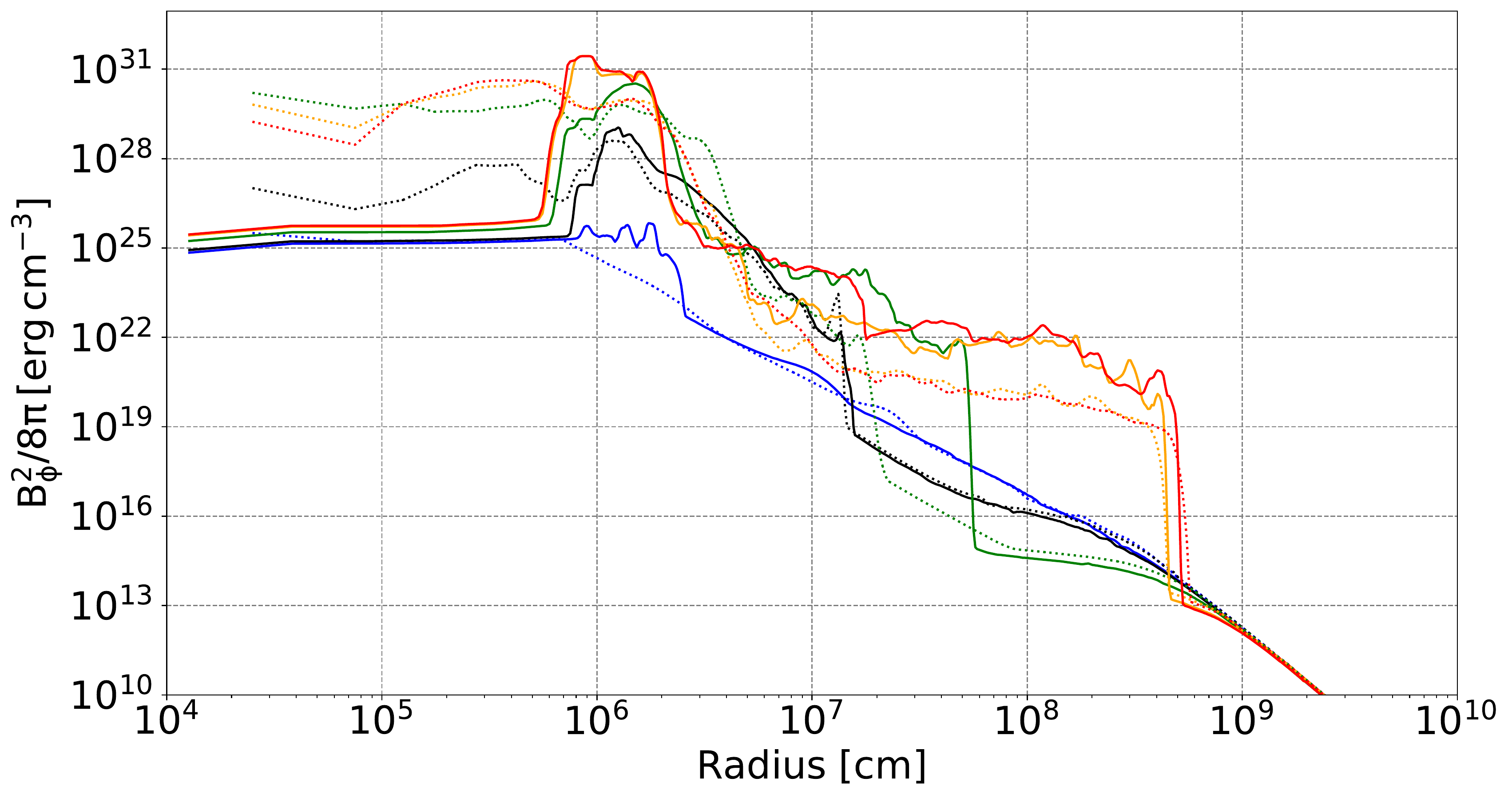}
\caption{Top to bottom:
Profiles of the energy density
contained in the radial, meridional,
and zonal components
$B_r$, $B_\theta$, and $B_\varphi$ 
of the magnetic field for the weak-field
models coco-HLLC-Rw (solid lines)
and  Alcar-Rw (dashed lines) at various post-bounce times.
}
\label{fig:MagWeakProfile}
\end{figure}

\begin{figure}
  \includegraphics[width=\linewidth]{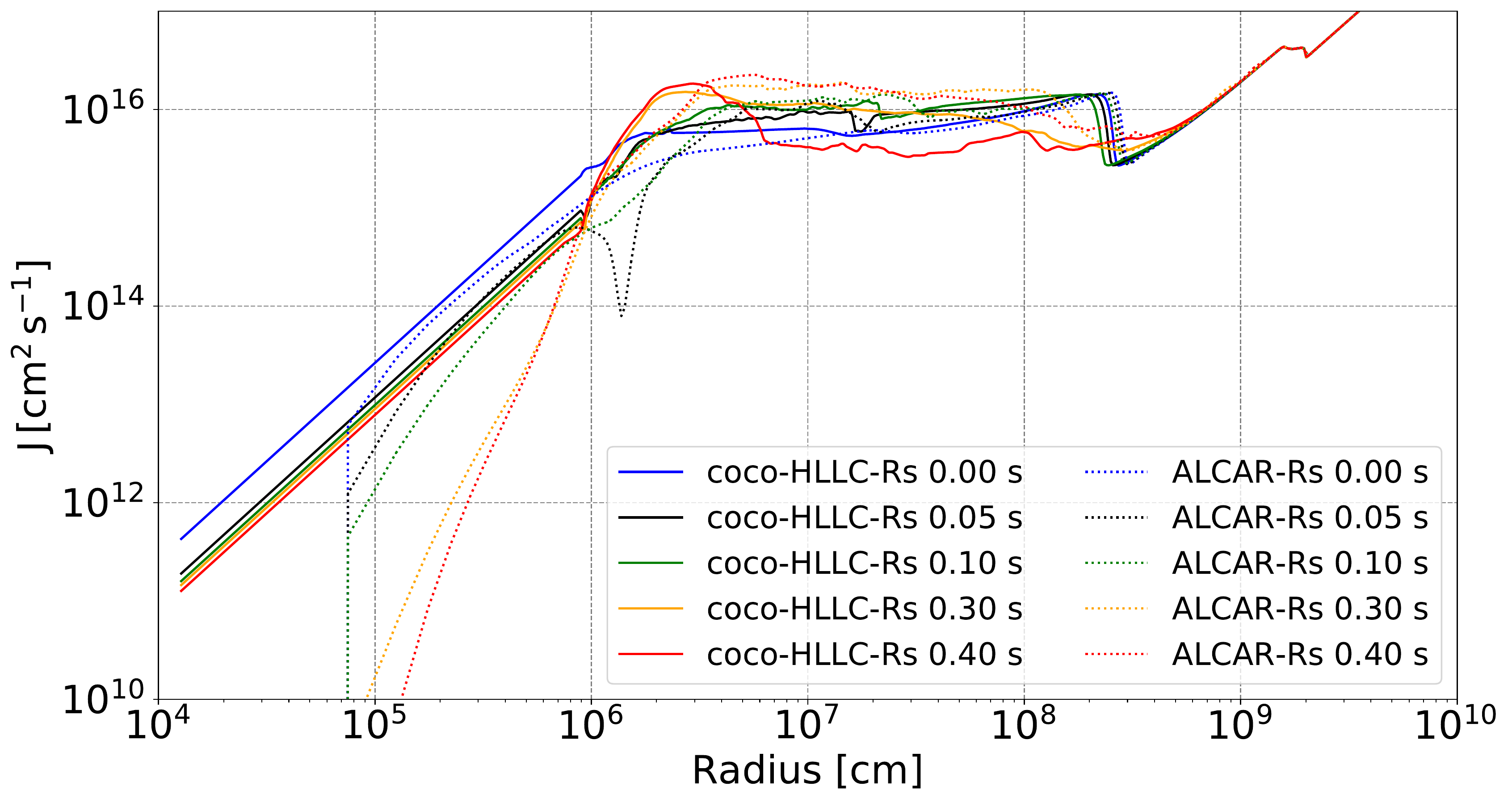}
  \caption{Spherically averaged specific angular momentum as a function of radius at different post-bounce times for the strong-field models Alcar-Rs
  (dotted lines) and coco-HLLC-Rs (solid lines).
  }
  \label{fig:StrongJ}
  \includegraphics[width=\linewidth]{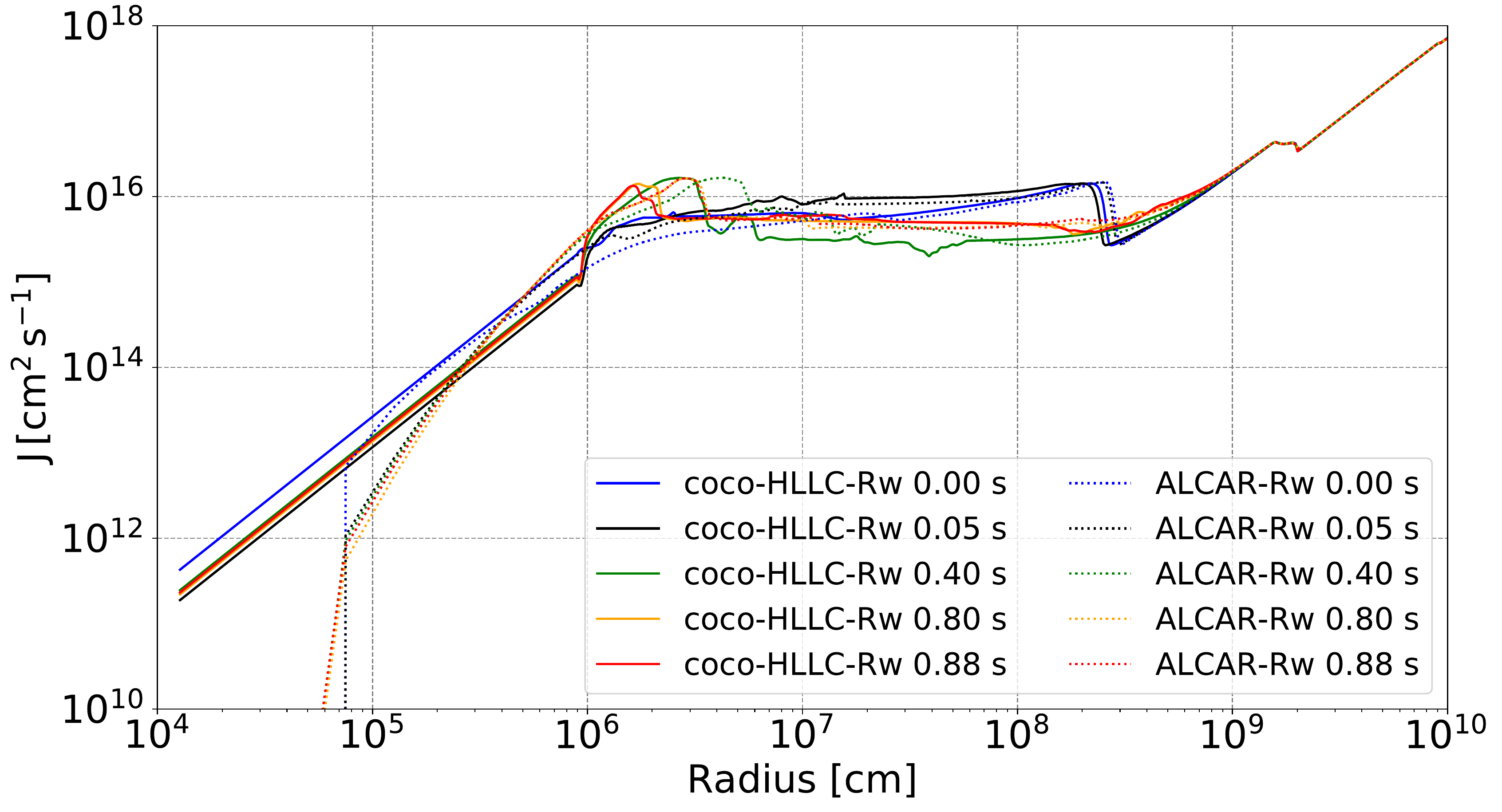}
  \caption{Spherically averaged specific angular momentum as a function of radius at different post-bounce times for the weak-field models Alcar-Rw
  (dotted lines) and coco-HLLC-Rw (solid lines).}
  \label{fig:WeakJ}
\end{figure}

The different evolution
of the magnetic field energy results from
a combination of effects, as revealed
by a closer analysis of radial profiles of the magnetic  field
energy density.
To this end we show spherical
averages of the components 
$e_{B,r}$,
$e_{B,\theta}$, and $e_{B,\varphi}$
of the magnetic
energy density contained in the radial, meridional,
and zonal components of the magnetic
field,
\begin{equation}
e_{B,r}=\frac{B_r^2}{8\pi},
\quad
e_{B,\theta}=\frac{B_\theta^2}{8\pi},
\quad
e_{B,\varphi}=\frac{B_\varphi^2}{8\pi},
\end{equation}
for the strong-field case
in Figure~\ref{fig:MagProfile} and for
the weak-field case in Figure~\ref{fig:MagWeakProfile}.
These profiles of the magnetic energy
are best viewed in conjunction with
angle-averaged profiles of the
specific angular momentum, which is
computed as
\begin{eqnarray}
j=
\frac{\int \rho v_\varphi r \sin \theta\,\ud\Omega}{\int \rho \,\ud\Omega}.
\end{eqnarray}
Profiles of $j$ at different times are
presented in Figure~\ref{fig:StrongJ}
and Figure~\ref{fig:WeakJ} for the
strong- and weak-field case, respectively.

Both for the weak- and strong-field case,
the profiles of 
$e_{B,r}$, $e_{B,\theta}$, and $e_{B,\varphi}$ 
are very similar at bounce, i.e., field
amplification during the collapse phase
in the inner core is consistent between the two codes despite the different treatment of the induction
equation. Later, however, the situation changes for two reasons. 

In the strong-field case, we 
first see a steady drop
of magnetic field energy in the innermost
few kilometres in the ACLAR-Rs run
because the inner core is treated in low-resolution 2D, leading to more numerical diffusion. In model coco-HLLC-Rs, strong
field dissipation in the core is avoided by
by enforcing uniform rotation and spherical
symmetry in the meridional flow components in the innermost $10\, \mathrm{km}$. 
Second, we see stronger field amplification
in model coco-HLLC-Rs around a radius
of $20\, \mathrm{km}$, primarily in 
$e_{B,\varphi}$ and to a lesser extent
in $e_{B,r}$, which is indicative of
strong shear amplification at the
PNS surface.  This is consistent with the finding
that the angular momentum profiles show a hump around $20\, \mathrm{km}$ in model coco-HLLC-Rs,
whereas the specific angular momentum
plateaus outside the PNS
than in Alcar-Rs (Figure~\ref{fig:StrongJ}).
This implies a steeper negative angular
velocity gradient, i.e., stronger shear,
in around and outside $20\, \mathrm{km}$
in coco-HLLC-Rs. The stronger
fields generated around these radii 
in coco-HLLC-Rs also
translate into stronger fields in the
jet in the long term; the components 
$e_{B,r}$ and $e_{B,\varphi}$ are about
a factor ten higher at the maximum shock
radius at late times. The profiles
of $e_{B,r}$ reveal a numerical disadvantage
of the divergence cleaning scheme as originally
implemented in \textsc{CoCoNuT-FMT}; choosing the
maximum cleaning speed allowed by 
the CFL limit results in spurious transport of magnetic energy into the pre-shock region.
The different treatment of the inner core
in both codes has further implications for the long-term evolution of magnetic fields in the PNS. While Alcar-Rs initially loses magnetic
field energy in the innermost zones,
it also shows stronger field amplification in $e_{B,\theta}$ just inside $10\, \mathrm{km}$ in contrast to CoCoNuT-HLLC-Rs. This is due to the inclusion of differential rotation of the inner core in the Alcar code, which allows for field amplification in this region that
is responsible for the rise in the
PNS magnetic energy $E_\mathrm{mag}$
after the very rapid decreases in magnetic energy in $e_{B,r}$, $e_{B,\theta}$, and $e_{B,\varphi}$ from numerical dissipation
in the innermost zones (cp.\ Figure~\ref{fig:PNS}).

The different evolution of the PNS
magnetic energy in the weak-field case
(with a faster rise in  \textsc{Aenus-Alcar} than in \textsc{CoCoNuT-FMT})
can also be understood as a consequence
of the different core treatment. Model
Alcar-Rw again shows field amplification
by differential rotation inside the core whereas CoCoNuT-HLLC-Rw only shows a slow
decline in $e_{B,r}$ and $e_{B,\theta}$
and a slow rise in $e_{B,\varphi}$ by compression inside the innermost $10 \, \mathrm{km}$. Interestingly, the (putatively numerical) decline
of the magnetic field energy in the innermost
zones is less pronounced in model Alcar-Rw than in model Alcar-Rs, suggesting that this phenomenon is very sensitive to the dynamics near the origin of the grid. Field amplification in the differentially rotating PNS core generates substantial toroidal fields,
which dominate the PNS magnetic energy; ultimately, the toroidal component of the magnetic energy density is increased relatively uniformly out to a radius of about $20\texttt{-}30\, \mathrm{km}$. By contrast,
field amplification in 
model coco-HLLC-Rw is confined to the PNS surface region outside the uniformly rotating core at radii of $10\texttt{-}20\, \mathrm{km}$. Similar to the strong-field case, the amplification of the toroidal field by shear in this region is stronger than in the
\textsc{Alcar} model. Due to the stronger field amplification at the PNS surface, the PNS magnetic energy in model CoCoNuT-HLLC-Rw eventually overtakes model Alcar-Rw at late times, overcompensating for the spurious lack of
field amplification inside the PNS.


The angular momentum profiles of models coco-HLLC-Rs and Alcar-Rs (Figure~\ref{fig:StrongJ}) and of models coco-HLLC-Rs and Alcar-Rs (Figure~\ref{fig:WeakJ})
yield more insights into the causes of the different patterns of field amplification in the
\textsc{CoCoNuT} and \textsc{Alcar} models.
The strong-field model coco-HLLC-Rs
shows the build-up of a hump in specific angular momentum outside the PNS centred at radii of 
$40\texttt{-}50\, \mathrm{km}$ in the late-time profiles. The presence of such a hump implies
considerable shear in and above the PNS surface region, which explains the strong magnetic fields generated in this layer in model coco-HLLC-Rs
(Figure~\ref{fig:MagProfile}). In model
Alcar-Rs, the hump is less pronounced, more
smeared out, and located further outside. In the early profiles, where the hump is yet absent, 
the transition to a plateau in specific angular still occurs further outside in model Alcar-Rs compared to coco-HLLC-Rs, implying less shear and less field amplification in the PNS surface region.
Further analysis reveals that the different
angular momentum distribution in the two runs
is related to the different PNS radii seen
in Figure~\ref{fig:PNS_rad}; effectively the
PNS has more fluffy, disk-like surface in model
Alcar-Rs. It is difficult to decide a priori which of the codes captures the structure of the disk-like surface region more accurately. We shall see in Section~\ref{ResAnalysis} that \textsc{CoCoNuT-FMT} also tends to produce more a more fluffy PNS surface at higher resolution. This   tentatively suggests that the more fluffy PNS surface structure with a flatter angular momentum distribution is more realistic.

The angular momentum profiles deeper in the
PNS also reveal an interesting phenomenon.
In model Alcar-Rs, there is a clear secular drop in angular momentum at radii  of less than $10\, \mathrm{km}$, whereas model coco-HLLC-Rs shows little changes in the
angular momentum profiles in this region.
One might be tempted to connect the slow-down
of PNS rotation with numerical diffusion
of angular momentum into the innermost zone,
where the angular momentum drops dramatically. However, the fact that the secular slow-down
of PNS rotation is much less pronounced
in model Alcar-Rw (Figure~\ref{fig:WeakJ}) suggests that numerical diffusion of angular momentum only plays a secondary role. Instead, the slow-down of core rotation in model Alcar-Rs appears to be connected to the action of magnetic stresses that initially build up  due to field winding and then become strong enough to make the inner core spin more slowly (akin to the winding and unwinding of a spring). This is reflected by the decreasing toroidal field strength in the innermost zones in model Alcar-Rs at late times (Figure~\ref{fig:MagProfile}, bottom panel), and the process of winding and unwinding also contributes to the long-period oscillation in the PNS magnetic energy in this model
(Figure~\ref{fig:PNS}, bottom panel). In the angular momentum profile at $0.05\, \mathrm{s}$, we find an interesting trace of such long-term torsional motions in the form of a dip in the specific angular momentum outside a radius of $10\, \mathrm{km}$ (Figure~\ref{fig:StrongJ}).
The long-period oscillations due to the winding and unwinding of the field are  absent in the \textsc{CoCoNuT-FMT} models because the inner core is forced to rotate uniformly.

The angular momentum profiles of the weak-field models Alcar-Rw and coco-HLLC-Rw are much more similar (Figure~\ref{fig:WeakJ}).
Both models develop humps in their angular momentum profiles with sharp angular momentum
gradients at the outer flank of the hump.
The humps are, however, located at different radii, which reflects the aforementioned tendency towards a more inflated PNS surface
structure in Alcar (cp.\ Figure~\ref{fig:PNS_rad}). The evolution of the angular momentum profile in the PNS core
is quite similar despite different  numerical
treatments for this region. Different from the strong-field model, magnetic stresses remain too weak to substantially affect the rotational evolution of the PNS core over the time scales of the simulation.
Model coco-HLLC-Rw has higher specific angular momentum at
radii less than about $3\, \mathrm{km}$, possibly avoiding a slight loss of angular momentum into the central zone in Alcar-Rw.
On the other hand, it does not capture the moderate amount of differential rotation in the PNS core due to its core treatment, while
Alcar-Rw maintains the profile of differential
rotation quite stably after an initial drop
of specific angular momentum in the innermost
zones in the first $50\, \mathrm{ms}$ after bounce.

\subsection{Neutrino Emission}

\label{neutrino_emission}
Even in a magnetorotational supernova, neutrinos still
play a major role by influencing the nucleosynthesis
conditions in the outflows
\citep[e.g.,][]{Janka2012_review}, and in some cases also by 
providing subsidiary energy input into the outflows. Since the fast-multi group transport scheme used in \textsc{CoCoNuT-FMT} \citep{Mueller2015} and the two-moment method implemented in \textsc{Aenus-Alcar}
\citep{Just2015} represent quite different approaches to multi-group neutrino transport, and since the codes do not use exactly the same set of neutrino interaction rates, differences in the neutrino emission between the two codes are expected. However, the impact of the transport treatment on the neutrino emission can hardly be disentangled from the effect of the PNS structure, which is sensitive to the multi-dimensional dynamics of the models as outlined in the previous section. In contrast to 1D code comparisons, it would be a futile undertaking to fully track down the cause for all the differences in the neutrino emission. We therefore compare the neutrino emission in the two codes at a rather descriptive level, only occasionally highlighting phenomena that can be traced either to differences in the neutrino transport or to the model dynamics with greater confidence.

\begin{figure}
    \centering
	\includegraphics[width=\linewidth]{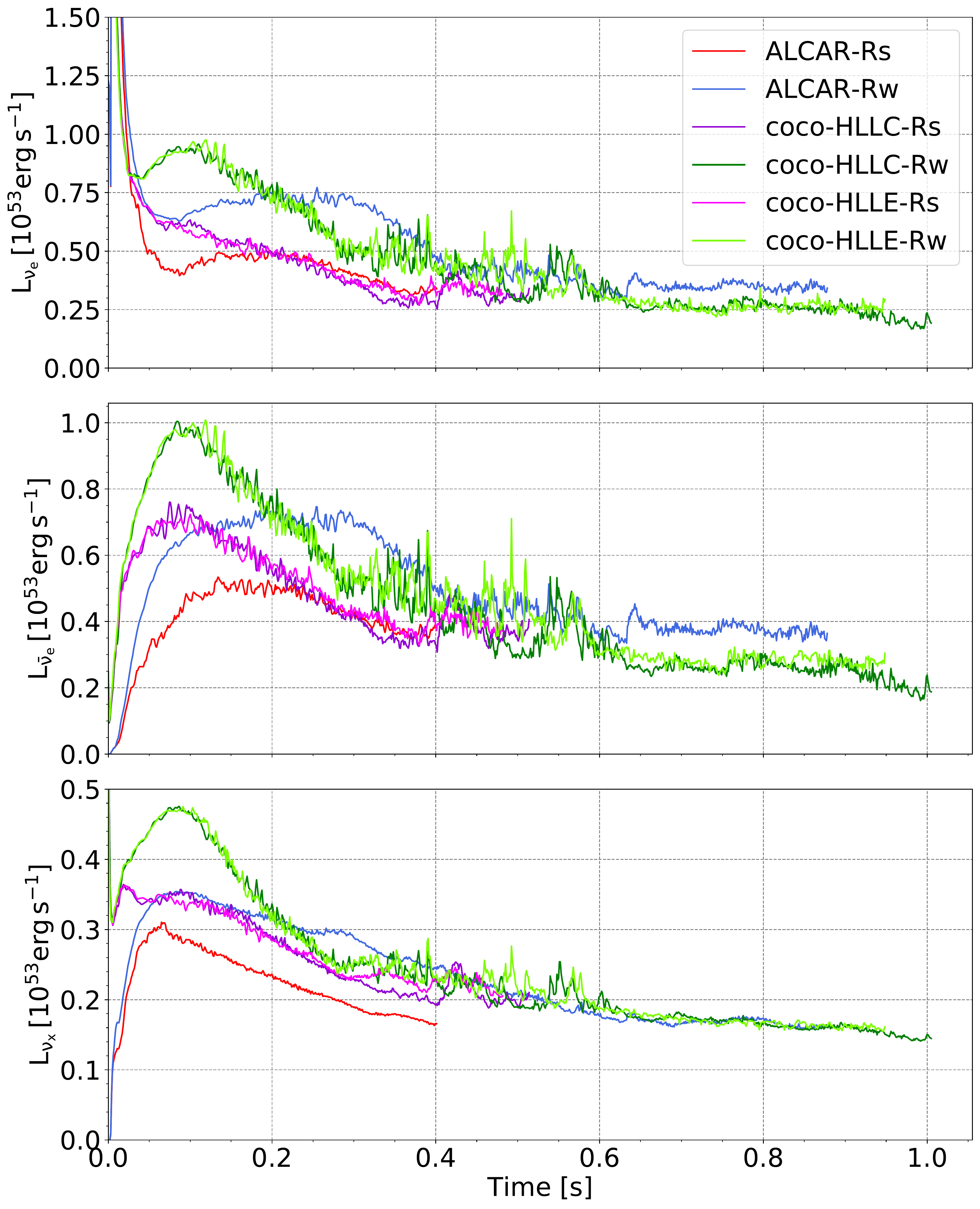}
\caption{Neutrino luminosities of
electron neutrinos ($\nu_\mathrm{e}$, top row), electron antineutrinos ($\bar{\nu}_\mathrm{e}$, middle) and  heavy-flavour neutrinos ($\nu_\mathrm{x}$, bottom), measured at a radius of $2000\,\mathrm{km}$.}
\label{fig:neutrinoLum}
\end{figure}

\begin{figure}
    \centering
	\includegraphics[width=\linewidth]{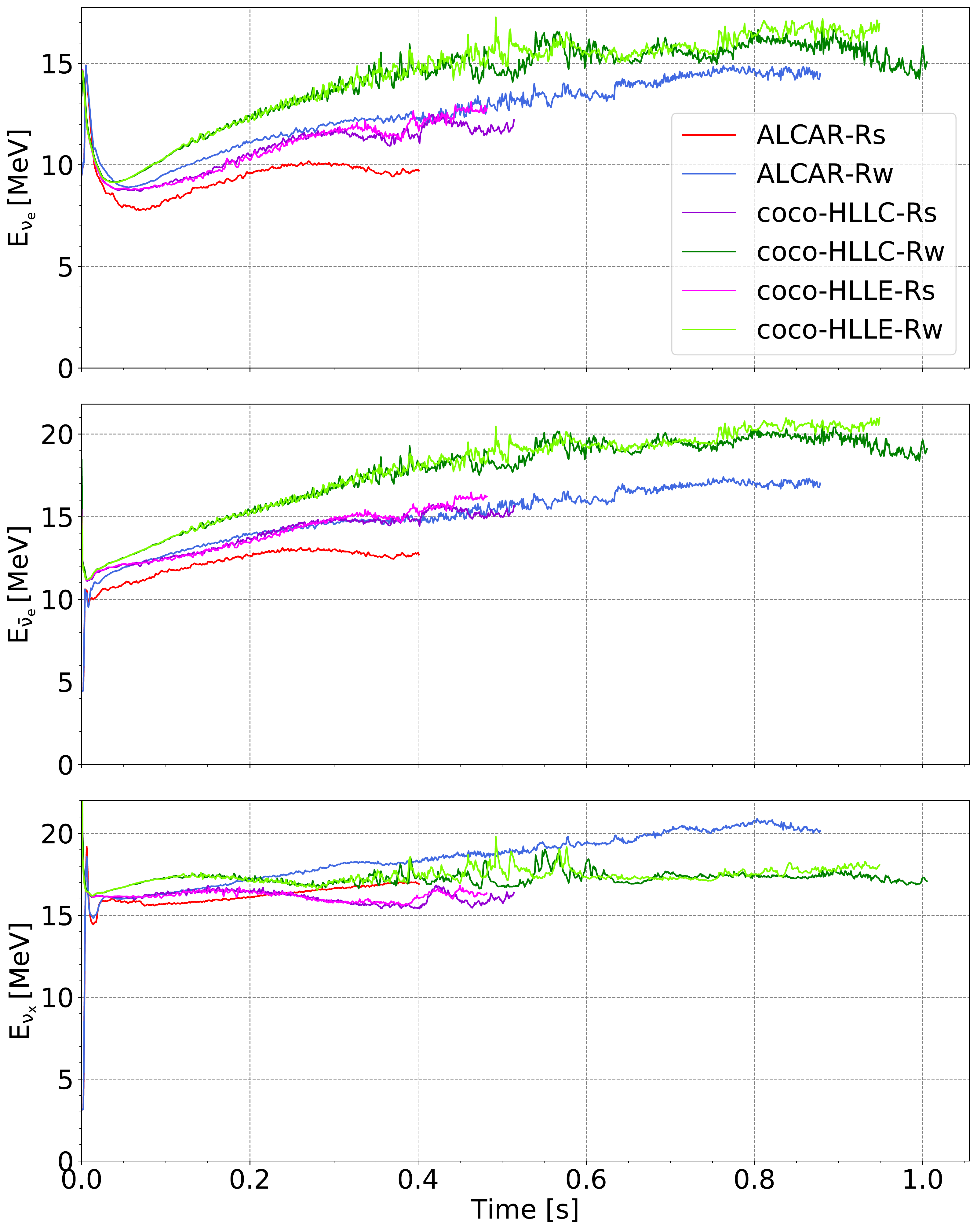}
\caption{Neutrino mean energies of
electron neutrinos ($\nu_\mathrm{e}$, top), electron antineutrinos ($\bar{\nu}_\mathrm{e}$, middle) and  heavy-flavour neutrinos ($\nu_\mathrm{x}$, bottom), measured at a radius of $2000\,\mathrm{km}$.}
\label{fig:neutrinoEne}
\end{figure}

Neutrino luminosities for the various \textsc{Alcar} models and the standard-resolution \textsc{CoCoNuT} model are shown in Figure~\ref{fig:neutrinoLum} and
reveal substantial differences.  Neutrino mean energies are shown in Figure~\ref{fig:neutrinoEne}.

Especially at early times, the neutrino luminosities of all flavours tend to be higher in the \textsc{CoCoNuT} models, while the corresponding \textsc{CoCoNuT} models with the
HLLC and HLLE solver show very similar luminosities. The higher early luminosities in \textsc{CoCoNuT} reflect differences in the collapse dynamics and early post-bounce accretion rate
(Figure~\ref{fig:accretion})
that were discussed in Section~\ref{shock_propagation}.

The temporal evolution of the neutrino mean energies is qualitatively similar, but \textsc{CoCoNuT} models are characterised by somewhat higher mean energies throughout, except for the heavy-flavour neutrinos.

In the case of the strong-field models, the 
electron neutrino and antineutrino luminosity
is higher by up to $2\times 10^{52}\,\mathrm{erg} \, \mathrm{s}^{-1}$ in models coco-HLLC-Rs and coco-HLLE-Rs compared to Alcar-Rs during the first $\mathord{\sim}100\, \mathrm{ms}$ after bounce and then gradually become more similar to the \textsc{Alcar} model. The higher electron-flavour mean energies are consistent with smaller PNS radii in \textsc{CoCoNuT}; the combination of smaller radii and higher surface temperatures fortuitously leads to similar neutrino luminosity as in \textsc{Alcar} after $\mathord{\sim}200\, \mathrm{ms}$. 
Note that unlike the electron-flavour luminosity, which is affected by differences
in the early post-bounce accretion rate, the heavy-flavour luminosity remains consistently higher in the strong-field \textsc{CoCoNuT} models throughout the simulation.  The faster contraction of the PNS in \textsc{CoCoNuT}  alone cannot account for the stronger emission since the heavy-flavour neutrino mean energy is not substantially higher in \textsc{CoCoNuT}. Different interaction rates likely play a substantial role here. The omission of electron-positron annihilation
in \textsc{CoCoNuT} can theoretically push the luminosity and mean energy up or down depending on the gradient of the temperature and chemical potentials near the neutrinosphere, the effective treatment of neutrino-nucleon scattering will tend to push the neutrino luminosity and mean energy down, and the strangeness corrections and nucleon correlations will likely push them upwards. 

It is noteworthy that there is a marked divergence between models coco-HLLC-Rs and coco-HLLE-Rs after $\mathord{\sim}0.4 \, \mathrm{s}$, which is when coco-HLLC-Rs develops a more fluffy PNS surface structure than coco-HLLE-Rs (Figure~\ref{fig:PNS}).
At this point, the electron-flavour mean energies in model coco-HLLC-Rs drop below
those in coco-HLLE-Rs by about $1\, \mathrm{MeV}$. Model coco-HLLC-Rs also
exhibits a noticeably smaller heavy-flavour luminosity than coco-HLLE-Rs from $0.3 \, \mathrm{s}$ onward. While the variations between these two models are considerably smaller than the differences between the \textsc{CoCoNuT} and \textsc{Alcar} models, this illustrates that
 neutrino luminosities and mean energies show considerable sensitivity to subtle dynamical variations in multi-dimensional MHD supernova models of rapidly rotating progenitors.

Among the weak-field models, we see similar differences in the early neutrino emission that can be ascribed to the same effects as for the strong-field models. In contrast to the strong-field models, there is, however, a cross-over in the electron flavour luminosities, with a phase of higher luminosities in 
Alcar-Rw compared to coco-HLLC-Rw and coco-HLLE-Rw. This is again consistent with the mass accretion rates in the two codes. At late times, the \textsc{CoCoNuT} models again exhibit higher electron-flavour luminosities; at this junction the more compact PNS surface structure is likely part of the explanation. Interestingly, the
heavy-flavour luminosities and mean energies become similar between \textsc{Alcar} and \textsc{CoCoNuT} at late times. As heavy-flavour neutrinos originate from higher densities, these are less affected by the different ``fluffiness'' of the PNS surface structure in the two codes.

The above results suggest that agreement 
between different codes in the neutrino luminosities and mean energies can only be achieved at a level of $10\texttt{-}20\%$ in MHD models of rapidly rotating supernova progenitors. This is decidedly worse than what was achieved in the comparison between the
\textsc{FMT} transport scheme and the \textsc{Vertex} neutrino transport code in
\citet{Mueller2015} in the non-rotating case.
The interplay between the multidimensional dynamics and the neutrino emission proves a considerable obstacle in analysing and eliminating code-dependent differences in the neutrino emission.

\subsection{Nucleosynthesis Conditions}
\label{nucleosynthesis}
\begin{figure}
  \includegraphics[width=\linewidth]{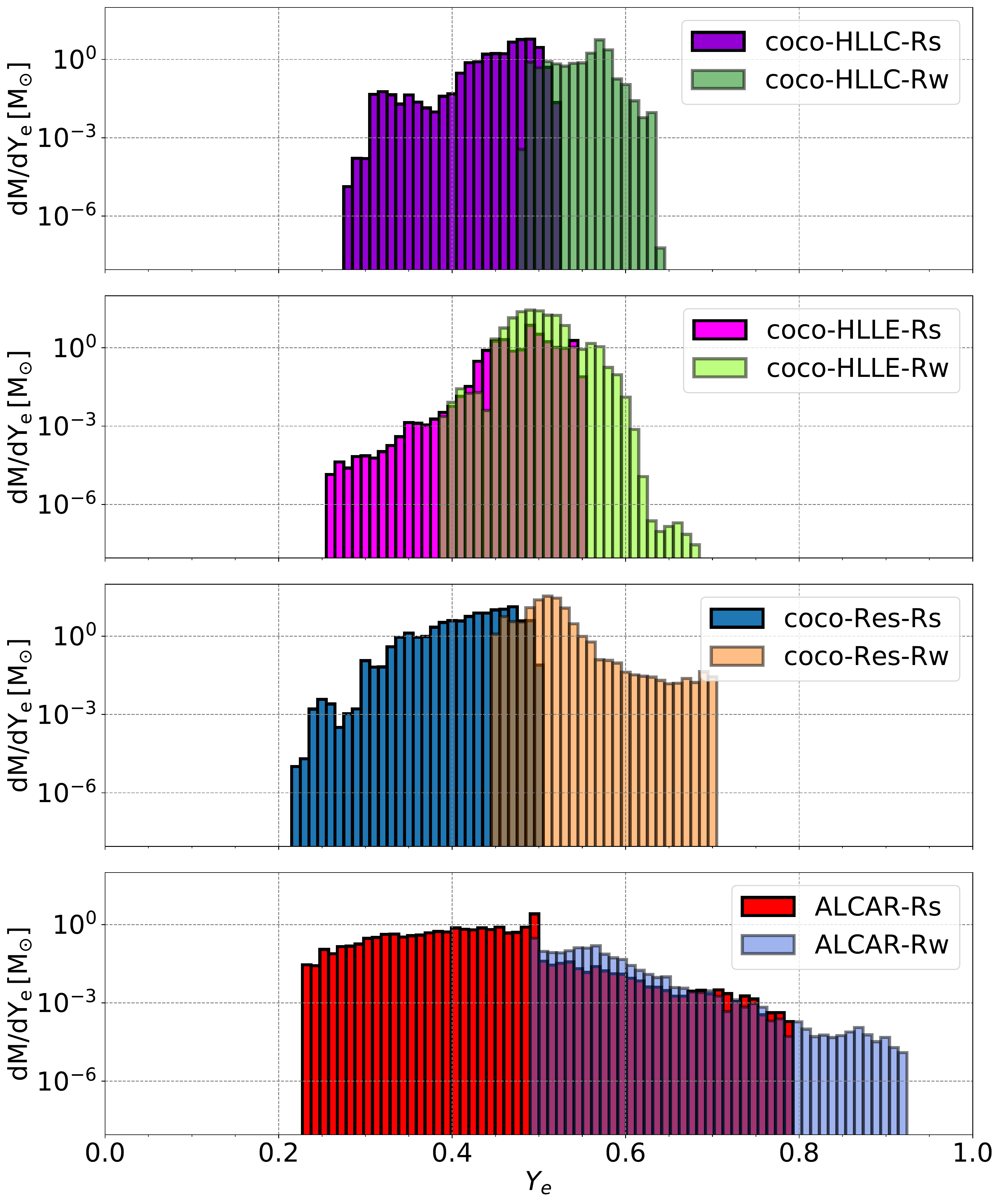}
  \caption{Histograms of the electron fraction $Y_e$ of the ejecta for the strong-field models at $\mathord{\approx} 0.40\,\mathrm{s}$ and at $\mathord{\approx} 0.85\,\mathrm{s}$ for the weak-field models. The four panels show, from top to bottom, the
  $Y_\mathrm{e}$-distribution in the coco-HLLC, coco-HLLE, coco-Res and Alcar models, respectively.
  }
  \label{fig:YeHist}
\end{figure}

Because of the ejection of neutron-rich material of relatively high entropy in the MHD-powered jets, magnetorotational supernovae have been proposed as an alternative site for the rapid neutron capture process 
\citep[r-process, e.g.,][]{Winteler2012,Halevi2018,Grimmett2021,Reichert2021}. This is supported by observational indications for r-process enrichment in low-metallicity environments that may be difficult to explain by neutron stars mergers \citep{Nishimura2015,Kobayashi2020}. In this context, it is important to analyse whether the different neutrino emission in \textsc{Alcar} and \textsc{CoCoNuT} has a substantial impact on the nucleosynthesis conditions in the outflows.

To this end, we visualise the distribution of the electron fraction $Y_\mathrm{e}$  in the ejecta in the different models using histograms for
all the weak- and strong-field models (Figure~\ref{fig:YeHist}).
The unbound ejecta are taken to comprise material with positive
total specific energy, positive radial velocity, and a density lower than
$10^{11}\, \mathrm{g}\, \mathrm{cm}^{-3}$ in order not to accidentally include material in the PNS.
Histograms are plotted at a post-bounce
time of $\mathrm{\approx} 0.4 \, \mathrm{s}$ for coco-HLLC-Rs, coco-HLLE-Rs, and Alcar-Rs, and at $\mathrm{\approx} 0.85\,\mathrm{s}$
for coco-HLLC-Rw, coco-HLLE-Rw, and Alcar-Rw.

All the strong-field models show a neutron-rich tail in the ejecta distribution that extends down to values
of $Y_\mathrm{e}$ as low as $0.25\texttt{-}0.3$, as
well as a smaller amount of proton-rich
ejecta. Neutron-rich conditions arise
in matter that is quickly ejected in the magnetically collimated jets from close to the PNS so that $Y_\mathrm{e}$ freezes out at low values before it can be raised by neutrino irradiation \citep{Wanajo2011,Muller_2016review}.
In slowly moving ejecta, neutrino absorption tends to establish proton-rich conditions because of similar $\nu_e$ and $\bar{\nu}_e$ luminosities and small differences between the $\nu_e$ and $\bar{\nu}_e$ mean energies \citep{FroehlichLiebendoerfer,Pruet}.
In all the weak-field models, which do not develop fast jets, the ejecta are predominantly proton-rich, similar to recent multi-D models of neutrino-driven explosion using multi-group
neutrino transport \citep{Wanajo2018,Eichler,Burrows2020},
although all models contain small amounts of neutron-rich ejecta except for Alcar-Rw.

Despite the differences in the neutrino-emission between \textsc{CoCoNuT} and \textsc{Alcar}, the qualitative shape of the $Y_\mathrm{e}$-distribution is quite similar in all models. This is not as surprising as it might first seem, since $Y_\mathrm{e}$ is less sensitive to the absolute values of the neutrino luminosities and mean energies than to the differences between electron neutrinos and antineutrinos, which are similar for both codes.

Nonetheless, we still find noticeable quantitative differences in the $Y_\mathrm{e}$-distribution. Once more, disentangling how the numerical methods and approximations employed in the two codes shape the distribution proves difficult, but a few plausible effects can be discerned.

Among the strong-field models,
coco-HLLE-Rs has the least amount of material with $Y_\mathrm{e}<0.4$. This is consistent with the more diffusive nature of the HLLE solver, which will tend to raise $Y_\mathrm{e}$ in the jets by numerical mixing with less neutron-rich material. However, the lowest $Y_\mathrm{e}$ in model coco-HLLE-Rs
is even \emph{lower} than in coco-HLLC-Rs, even though the total amount of material below $Y_\mathrm{e}<0.4$ is substantially smaller. This is not a contradiction because the HLLE solver sometimes produces more violent, unsteady flow with bigger flow structures in the subsonic regime \citep{MuellerPHD}.
The amount of strongly neutron-rich material in Alcar-Rs is higher than in the \textsc{CoCoNuT} models and $Y_\mathrm{e}$ reaches lower minimum values
($0.23$ in Alcar-Rs as opposed to $0.28$ in coco-HLLC-Rs). However, it is not
possible to disentangle whether this
is due to less severe numerical dissipation, different jet dynamics,
or the neutrino transport treatment.
The neutrino transport likely plays some role because model Alcar-Rs shows a rather remarkable proton-rich tail up to $Y_\mathrm{e}=0.8$, whereas the
strong-field \textsc{CoCoNuT} models only eject moderately proton-rich material. It is unlikely that numerical diffusion alone could compress the distribution
of proton-rich material so significantly as to explain the narrow distribution in the \textsc{CoCoNuT} model on its own.

The $Y_\mathrm{e}$-distributions in the weak-field models agree inasmuch as the bulk of the ejecta fall into the moderately proton-rich range of
$Y_\mathrm{e}=0.5\texttt{-}0.6$.
However, one notices a more extended proton-rich tail  in 
Alcar-Rw compared to the \textsc{CoCoNuT-FMT} models.
Unlike model Alcar-Rw, the \textsc{CoCoNuT} models also have a small admixture of slightly neutron-rich ejecta. Interestingly, the $Y_\mathrm{e}$-distribution in model
coco-HLLE-Rw is broader than in coco-HLLC-Rw and shifted to slightly smaller values. From the viewpoint of nucleosynthesis, the impact of the Riemann solver appears quite substantial. A shift of the minimum $Y_\mathrm{e}$ from $0.48$ in coco-HLLC-Rw to about $0.39$ in coco-HLLE-Rw will imply a considerable change in iron-group nucleosynthesis from NSE freeze-out from a predominant contribution of ${}^{56}\mathrm{Ni}$ to a substantial contribution of neutron-rich nuclides like $^{60}\,\mathrm{Fe}$ and $^{48}\,\mathrm{Ca}$ \citep{Wanajo2018,Hartmann1985} and others.
In summary, our code comparison so far
already suggests that yields from the tail of the $Y_\mathrm{e}$-distribution in MHD supernova models are expected to be quite sensitive to numerical details, whereas the composition of the bulk of the ejecta close to $Y_\mathrm{e}=0.5$ appears more robust. This suggests that
the models can be more confidently used to predict ${}^{56}\mathrm{Ni}$ yields that are relevant for supernova light curves, as ${}^{56}\mathrm{Ni}$ is produced in the regime close to and above $Y_\mathrm{e}=0.5$, whereas r-process yields from simulations may be prone to considerable uncertainties.

\section{Results -- Resolution Study}\label{ResAnalysis}
To supplement the code comparison in Section~\ref{models},  we compare models coco-HLLC-Rs and coco-HLLC-Rw to the corresponding high-resolution models coco-Res-Rs and coco-Res-Rw, which have
been calculated with an angular resolution of $0.7^\circ$
(256 zones) instead of $1.4^\circ$ (128 zones). By revisiting parts of the analysis in Section~\ref{results} we shall not only explore a further dimension of uncertainties in MHD supernova models, but also obtain clues about the reasons underlying the differences between \textsc{Alcar} and \textsc{CoCoNuT} models.

\begin{figure}
  \includegraphics[width=\linewidth]{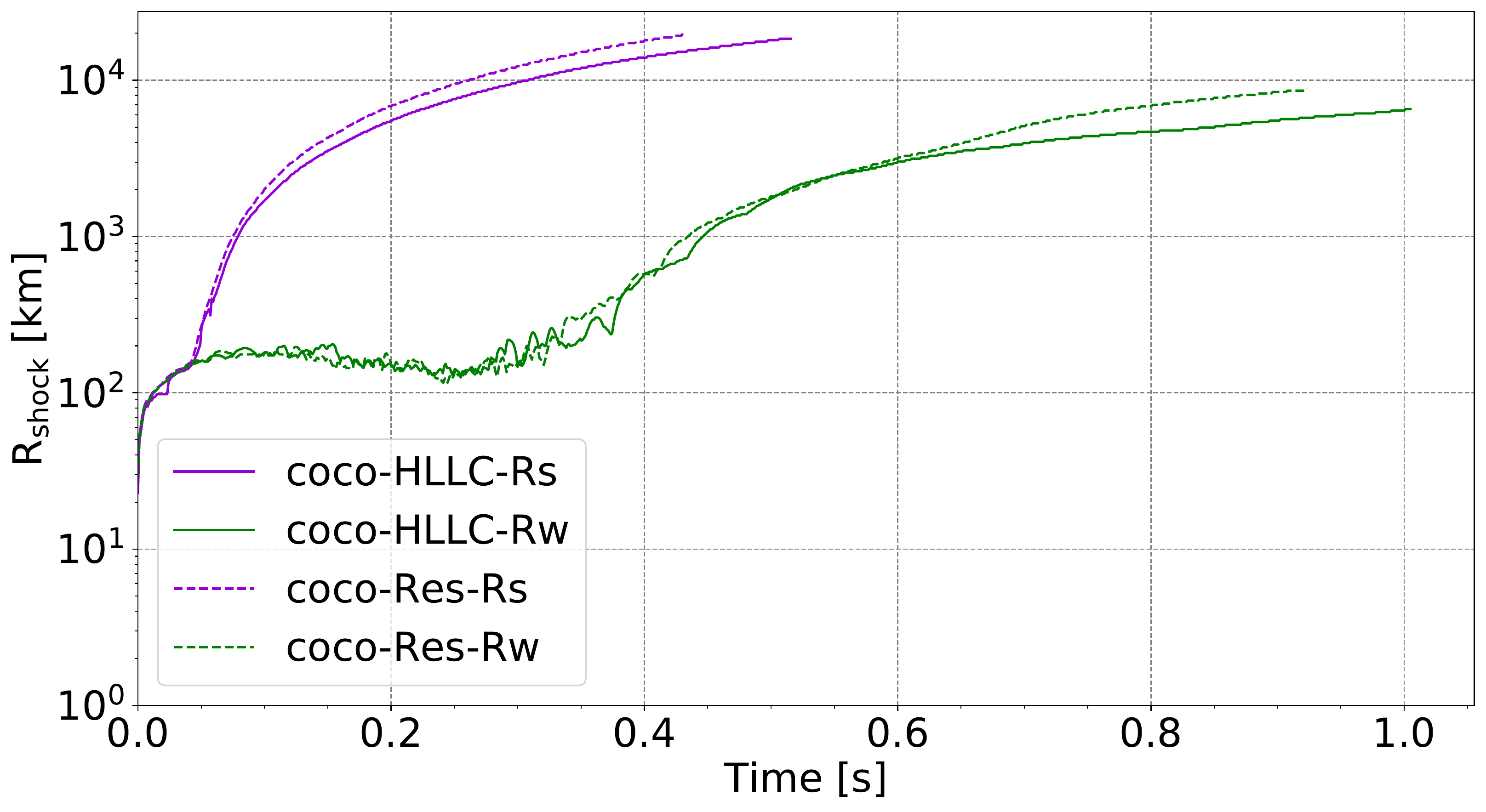}
  \caption{Evolution of the maximum shock radius for the coco-HLLC models
  at standard resolution compared to the corresponding high-resolution coco-Res models.
  Both the weak- and strong-field case are shown.
  field cases. The radius given in a log10 scale.}
  \label{fig:ShockRadiusRes}
\end{figure}
 
 \subsection{Shock Propagation and Explosion Energy}
\begin{figure}
    \includegraphics[width=\linewidth]{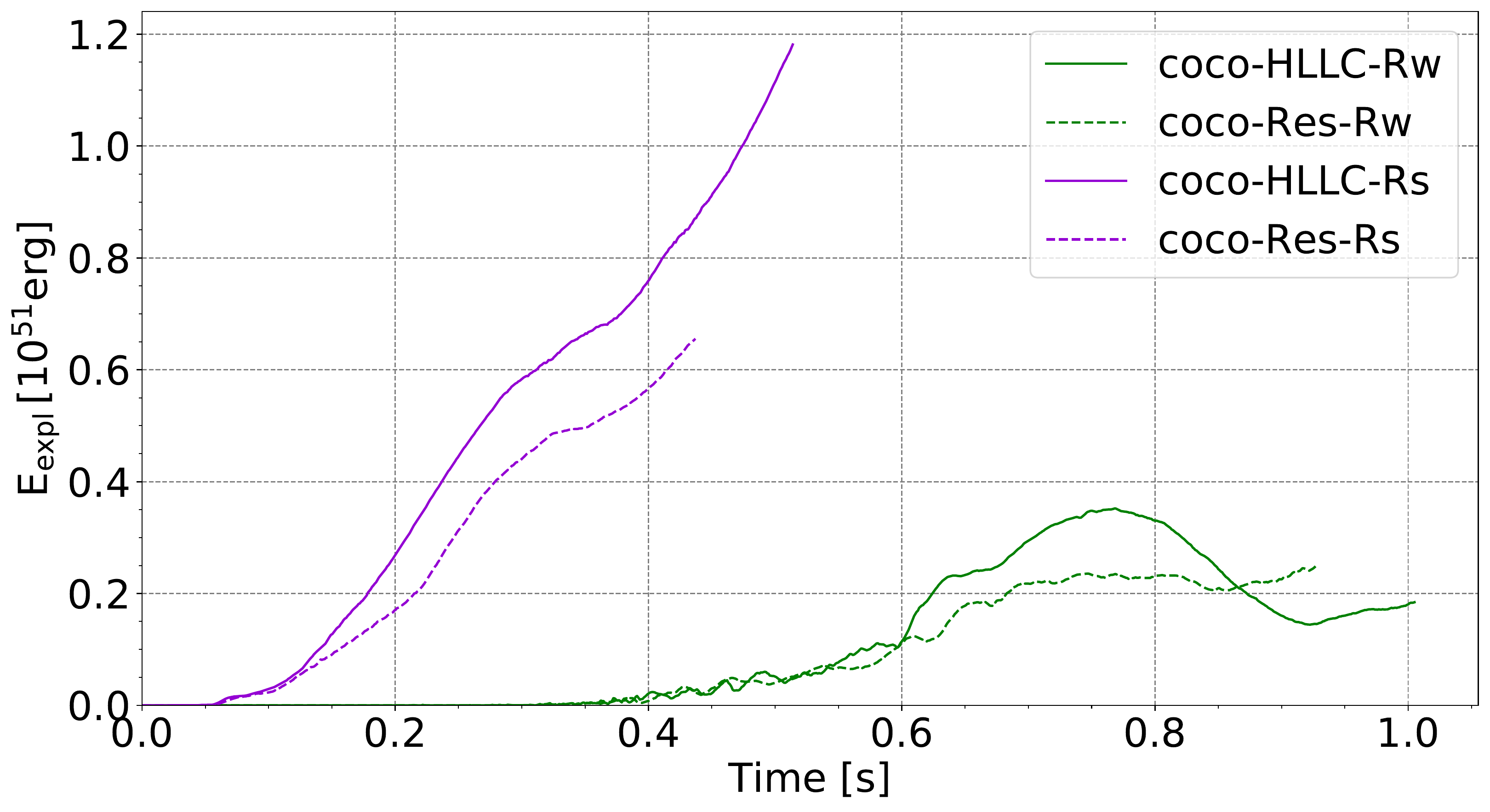}
    \caption{Comparison of the evolution of the diagnostic explosion energy
    $E_\mathrm{expl}$ in the strong- and weak-field models coco-HLLC
    at standard resolution to the corresponding high-resolution
    coco-Res models.}
    \label{fig:ExplosionEneRes}
\end{figure}

Figures~\ref{fig:ShockRadiusRes} and \ref{fig:ExplosionEneRes} show the shock trajectories and explosion energies of the standard- and high-resolution models as the simplest global metrics of the dynamics. Among the strong-field models, we note that the high-resolution model coco-Res-Rs exhibits somewhat faster shock expansion than the baseline model coco-HLLC-Rs. The explosion energy is lower in  coco-Res-Rs, however, and grows only at a rate of $\mathord{\approx} 1.7 \times 10^{51}\, \mathrm{erg}\,\mathrm{s}^{-1}$ as opposed to $\mathord{\approx} 2 \times 10^{51}\, \mathrm{erg},\mathrm{s}^{-1}$ in coco-HLLC-Rs. This suggests that uncertainties of order $\mathord{\sim}10\%$ in these global quantities are expected due to limitations in resolution, or possibly from stochastic model variations \citep[cp.][]{Cardall2015a} alone.
We note that the small explosion energy of the high-resolution model is not at odds with the resolution study of  \citet{Sawai2015}, who found a faster growth of the explosion energy for higher resolution in MHD-driven supernova models, but considered a somewhat different scenario. \citet{Sawai2015} started their simulations with smaller seed fields and attempted to follow field amplification by the MRI using very high resolution \emph{inside} the PNS, whereas our simulations start with sufficiently strong fields to promptly trigger a magnetorotational explosion.

For the weak-field models, we observe small differences in the shock propagation and explosion energy between models coco-Res-Rw and coco-HLLC-Rw as well.
Towards the end of the simulations, coco-Res-Rw has a larger maximum shock radius, but the difference in shock propagation is quite variable over time. Similarly, coco-Res-Rw has a lower explosion energy until $0.9\, \mathrm{s}$ but then overtakes model coco-HLLC-Rw. The variations between these two models may thus be purely stochastic in nature.

 \subsection{Proto-neutron Star Structure and Neutrino Emission}
\begin{figure}
  \includegraphics[width=\linewidth]{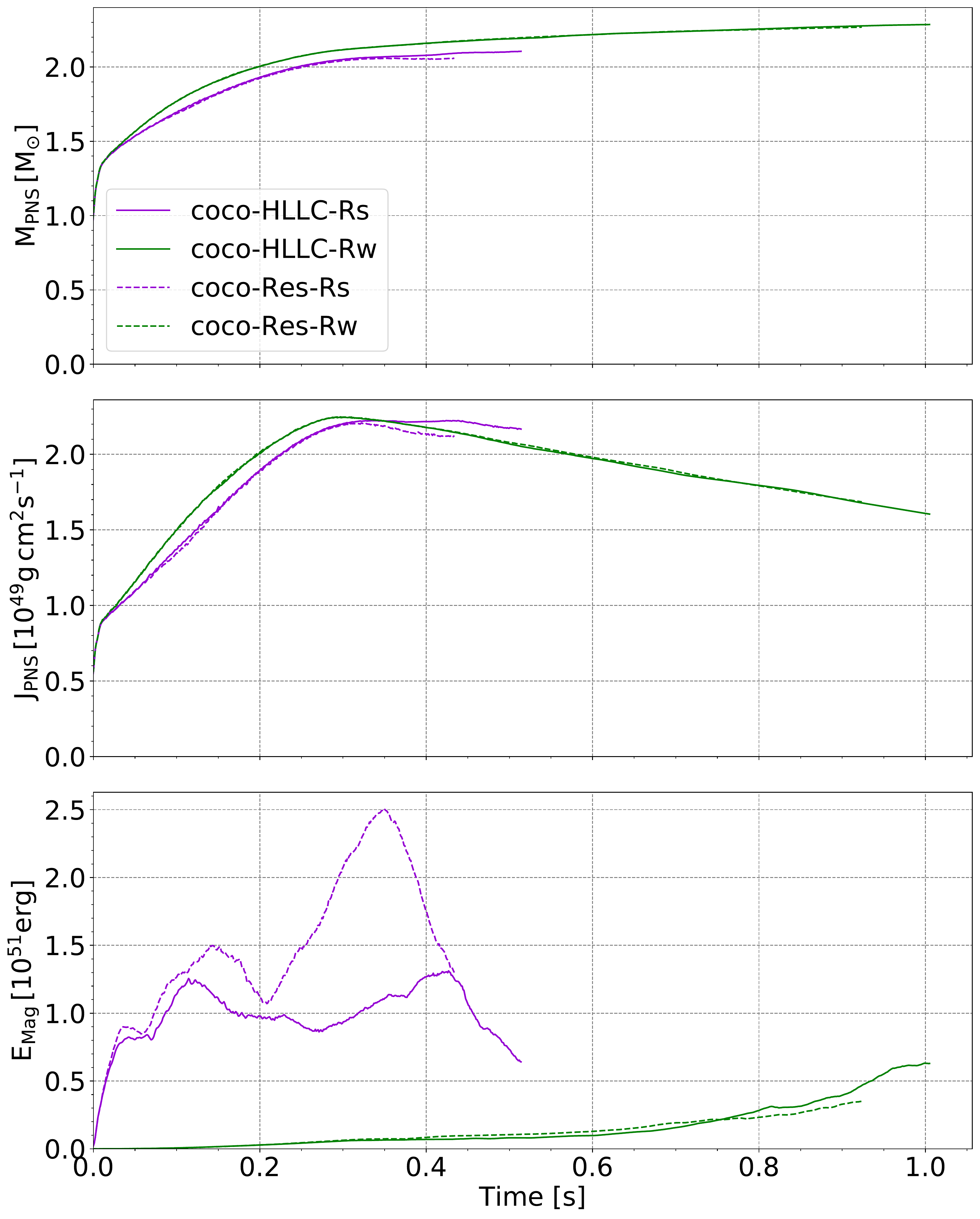}
\caption{Comparison of the PNS mass
  $M_\mathrm{PNS}$, angular momentum
  $J_\mathrm{PNS}$, and magnetic energy 
  $E_\mathrm{mag}$,
  as a function of post-bounce time for the standard-resolution 
  coco-HLLC models and the high-resolution coco-Res models.   The PNS is defined as
  the region where $\rho > 10^{11} \,\mathrm{g}\,\mathrm{cm}^{-3}$.
  }
    \label{fig:PNS_ValuesRes}
\end{figure}

\begin{figure}
    \includegraphics[width=\linewidth]{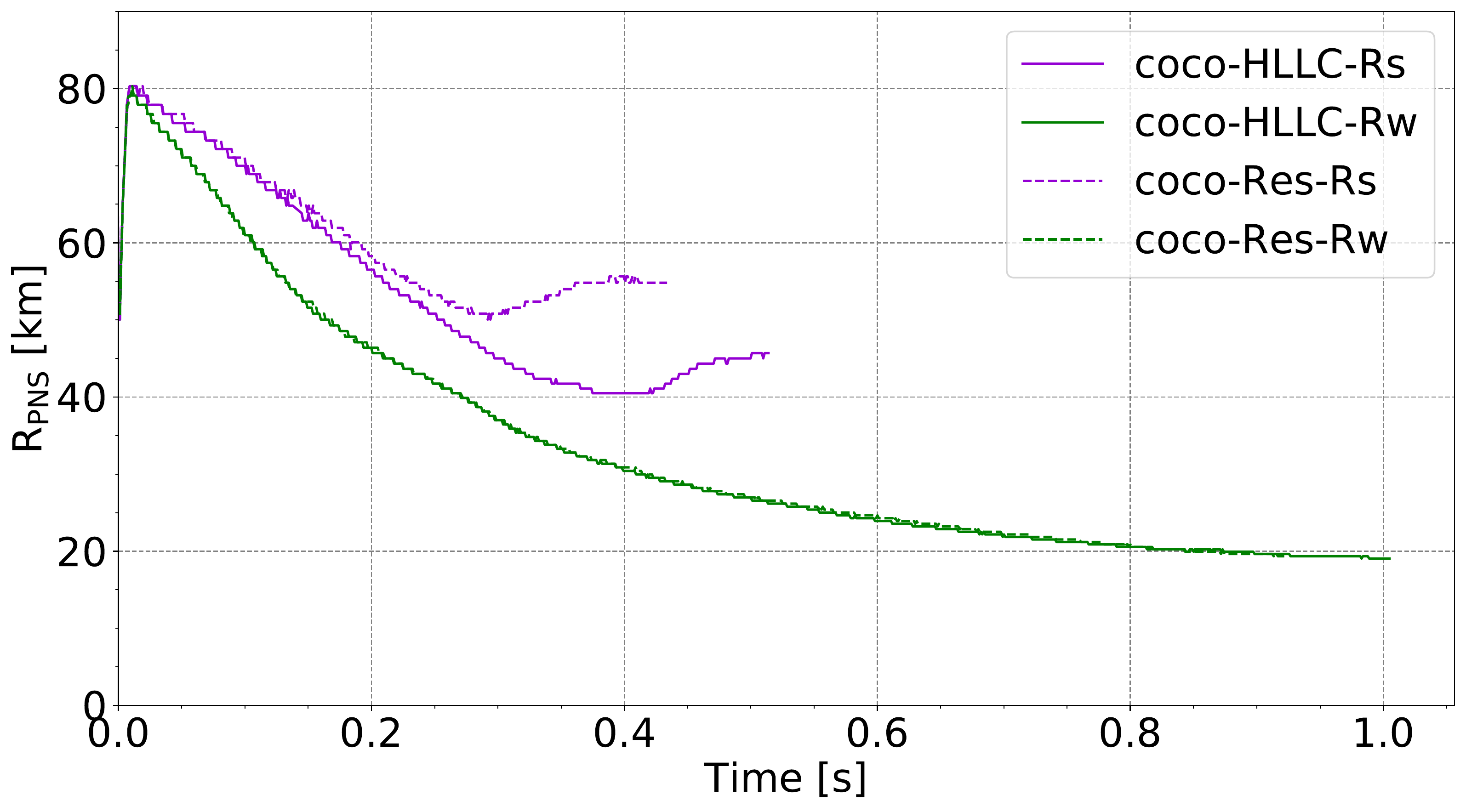}
  \caption{Comparison of the time evolution of the angle-averaged PNS radius
  $R_\mathrm{PNS}$ 
  between \textsc{CoCoNuT-FMT} models with different resolution.}
  \label{fig:PNS_radRes}
\end{figure}

 Figure~\ref{fig:PNS_ValuesRes} shows
 the PNS mass, angular momentum, and magnetic energy for models coco-Res-Rs,
 coco-Res-Rw, coco-HLLC-Rs, and coco-HLLC-Rw. In the weak-field case, the high-resolution and standard-resolution runs show almost perfect agreement in these quantities. There is only a hint of a slightly faster rise of the PNS magnetic energy
 in coco-HLLC-Rw close to the end of the simulation. However, more substantial differences emerge between the strong-field models coco-HLLC-Rs and coco-Res-Rs. After about $0.3 \, \mathrm{s}$, the PNS angular momentum in model coco-Res-Rs starts to decrease faster than in coco-HLLC-Rs and ends up with a $5\%$ smaller value. More intriguingly,  the PNS magnetic energy in model coco-Res-Rs  peaks at almost double the value in coco-HLLC-Rs.
The evolution of the PNS radius also differs significantly (Figure~\ref{fig:PNS_radRes}). Model coco-Res-Rs shows significant radius inflation at about $0.3\, \mathrm{s}$, and while a similar phenomenon occurs in the standard-resolution run, this happens about $0.1 \, \mathrm{s}$ later. This is noteworthy because a similar divergence of the PNS radii was found in our comparison between the strong-field \textsc{Alcar} and \textsc{CoCoNuT} models in Section~\ref{results}.
This is a further indication that the PNS surface region is subject to some form of tipping process that results in the adjustment to a more stable structure under conditions that are highly sensitive to the numerical implementation and grid resolution. Models coco-HLLC-Rs and coco-Res-Rs offer an opportunity to visualise the compact and inflated state of the PNS more clearly.

\begin{figure}
  \centering
  \includegraphics[width=\linewidth]{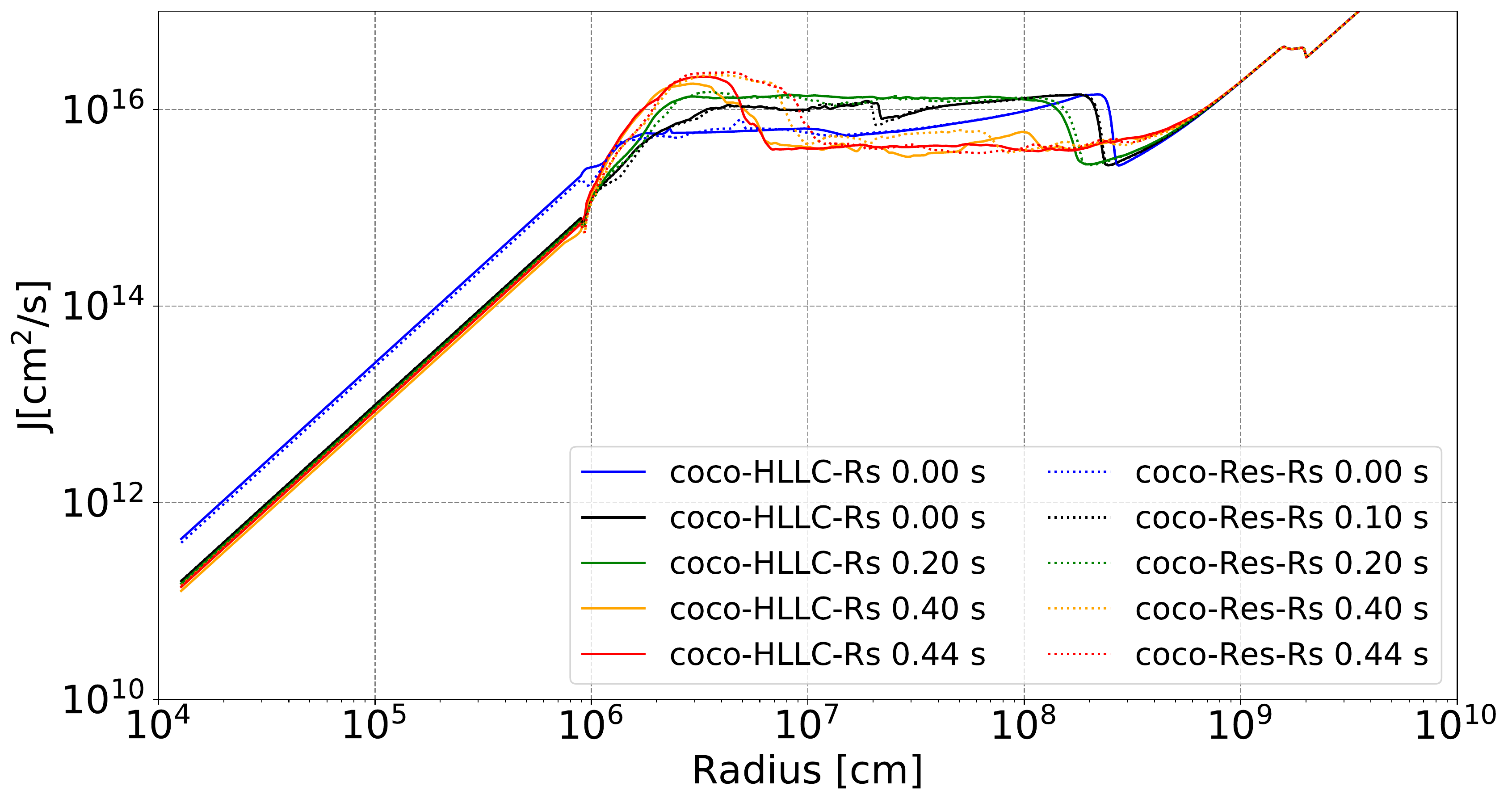}
  \caption{Spherically averaged specific angular momentum as a function of radius at different post-bounce times for the strong-field models coco-Res-Rs
  (dotted lines) and coco-HLLC-Rs (solid lines).
  }
  \label{fig:StrongJRes}
\end{figure}

\begin{figure*}
    \centering
    \includegraphics[width=0.49\linewidth]{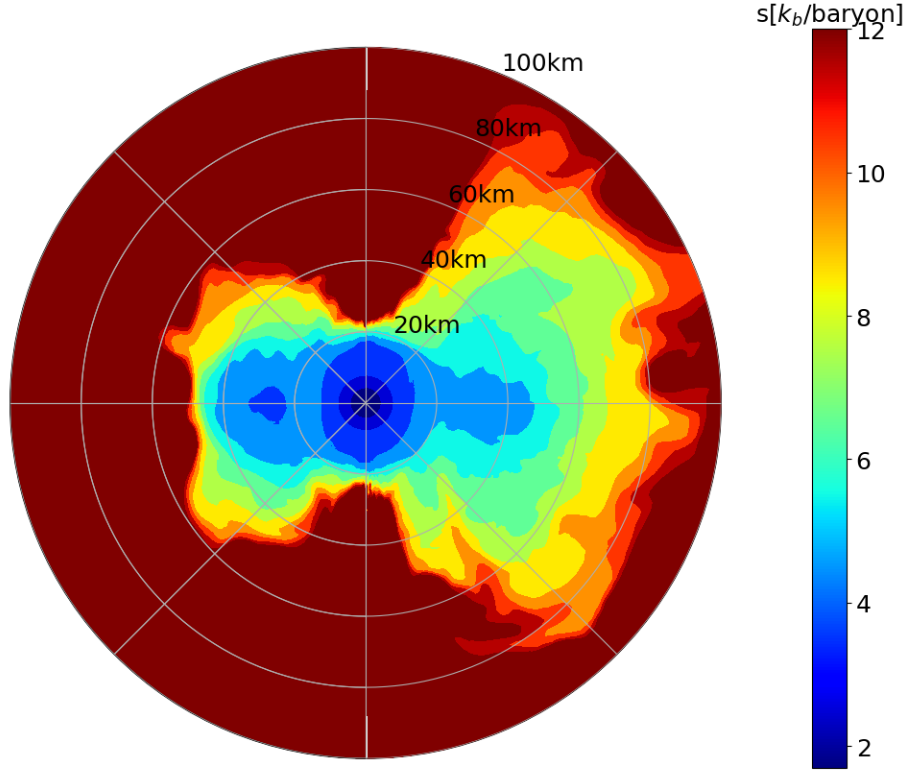}\hfill
    \includegraphics[width=0.49\linewidth]{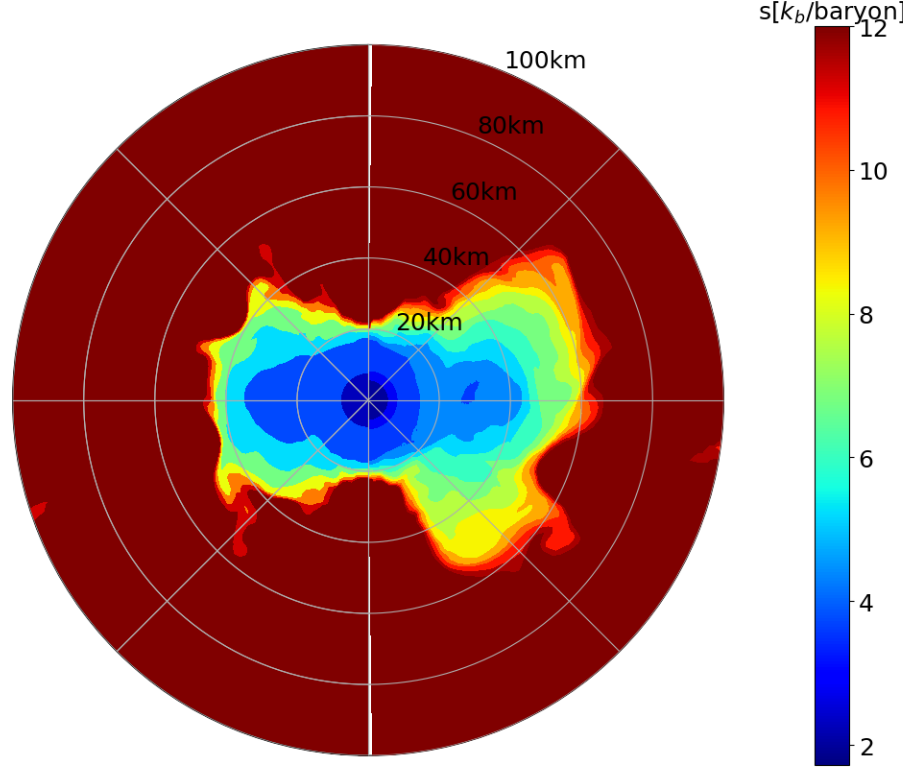}
    \caption{Left: 2D plots of the entropy $s$ in the vicinity of the PNS
    for the standard-resolution model coco-HLLC-Rs (left half) and the
    high-resolution model coco-Res-Rs (right half) at $\mathord{\approx} 0.46\, \mathrm{s}$.  Right: Corresponding plot of the entropy $s$ in model
    coco-HLLE-Rs (left half) and coco-HLLC-Rs (right half)
    at $\mathord{\approx} 0.45\,\mathrm{s}$
    The entropy in both plots is capped at $12 k_\mathrm{b}/\mathrm{baryon}$ to make
    the low-entropy, disk-like PNS surface region more clearly visible.}
       \label{fig:PNS_entropyCoCo}
\end{figure*}

In Figure~\ref{fig:PNS_entropyCoCo} (left panel), we show 
2D plots of the entropy in the innermost $100\, \mathrm{km}$ of the grid in models coco-Res-Rs
and coco-HLLC-Rs. In both cases, the PNS has a strongly oblate structure, and its surface even forms something of a torus or thick disk. In the high-resolution model coco-Res-Rs, the disk-like structure is considerably more extended than in model coco-HLLC-Rs. We note that the edge of the PNS accretion disk at about $50\, \mathrm{km}$ in  coco-HLLC-Rs corresponds to the steep gradient in angular momentum profile at late times in this model
(Figure~\ref{fig:StrongJRes}, cp.\ also
Figure~\ref{fig:StrongJ}).
This suggests that the inflation of the disk is caused by outward angular momentum transport from the inner regions of the PNS by strong magnetic fields to larger radii. This  is consistent with our interpretation that the different angular momentum profiles in the \textsc{Alcar} and \textsc{CoCoNuT} discussed in Section~\ref{results} are due a bifurcation instability that leads to a different PNS surface structure when triggered.

We can see a similar effect, albeit at smaller level,
in the comparison between models coco-HLLE-Rs and coco-HLLC-Rs
with the latter exhibiting a more fluffy disk structure  than
the HLLE model (Figure~\ref{fig:PNS_entropyCoCo}, right panel). Intriguingly, the more dissipative Riemann solver produces a disk structure that is less inflated than with the more accurate HLLC solver. Less numerical dissipation on the grid scale -- either
due to the use of a more accurate solver or higher resolution -- seems to enhance rather than to inhibit angular momentum transport. This appears counterintuitive at first glance, but there are other known instances of instabilities being \emph{suppressed} by insufficient numerical dissipation, e.g., spurious behaviour of the Kelvin-Helmholtz instability in smoothed-particle hydrodynamics can be cured
by introducing appropriate artificial conductivity at contact discontinuities \citep{Price2008}. On the other hand, the instability could still arise more easily at higher resolution due to increased small-scale numerical noise. Alternatively, the instability might be triggered by stronger magnetic field amplification in model coco-Res-Rs (Figure~\ref{fig:PNS_ValuesRes}, bottom panel).

Regardless of the reason, the divergent disk structures in models coco-HLLC-Rs, coco-Res-Rs, and coco-HLLE-Rs further illustrate that the dynamics of MHD supernova models is highly sensitive to details of the numerical implementation such as the Riemann solver, and to the grid resolution, even though some bulk metrics like the explosion energy do not immediately show the same strong dependency.

\begin{figure}
    \centering
	\includegraphics[width=\linewidth]{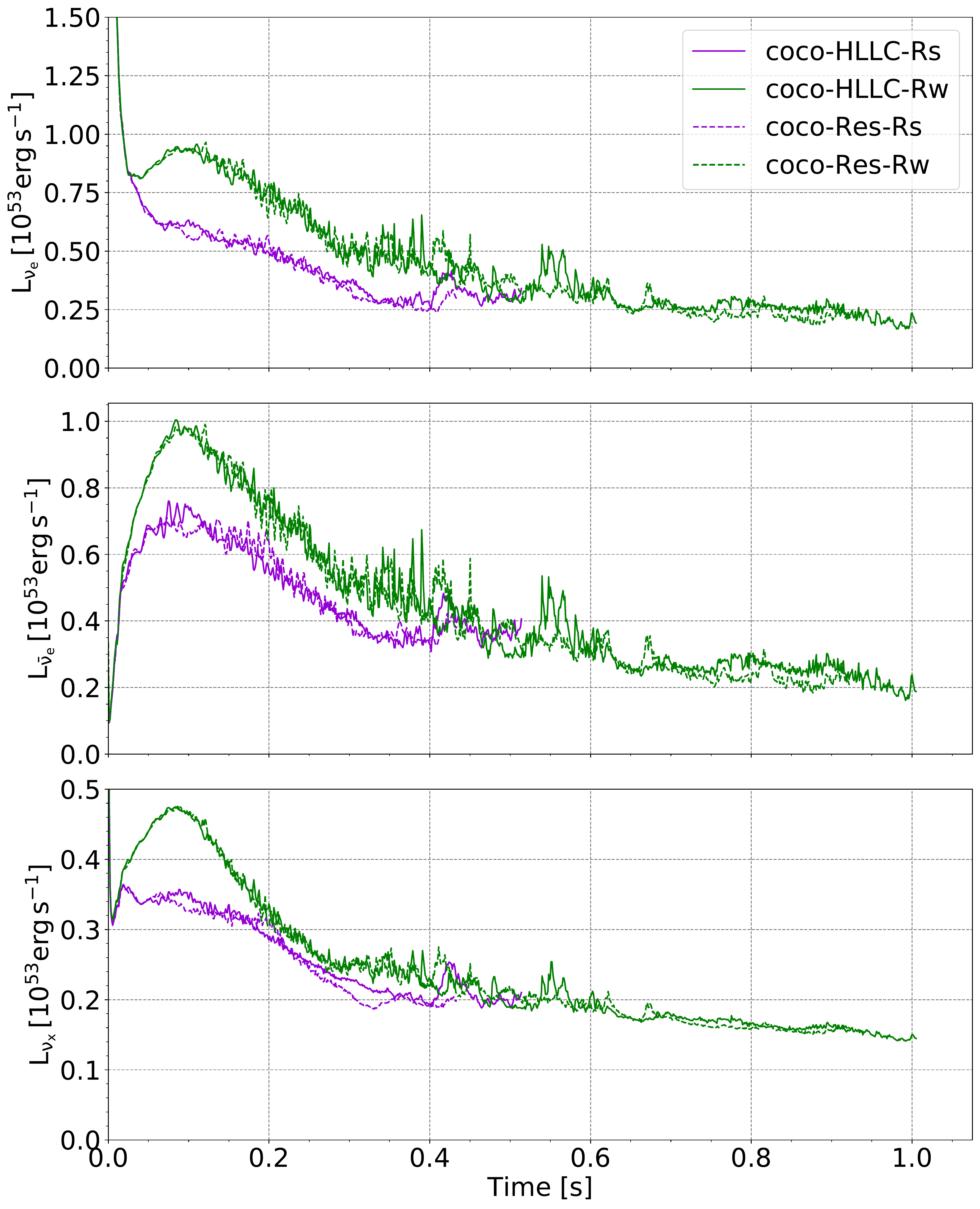}
\caption{Comparison of the luminosities of
electron neutrinos ($\nu_\mathrm{e}$, top row), electron antineutrinos ($\bar{\nu}_\mathrm{e}$, middle) and  heavy-flavour neutrinos ($\nu_\mathrm{x}$, bottom)
for the standard-resolution coco-HLLC models and the high-resolution coco-Res models.
Luminosities are measured as spherical averages at a radius of $2000\,\mathrm{km}$.}
\label{fig:neutrinoLumRes}
\end{figure}

\begin{figure}
    \centering
    \includegraphics[width=\linewidth]{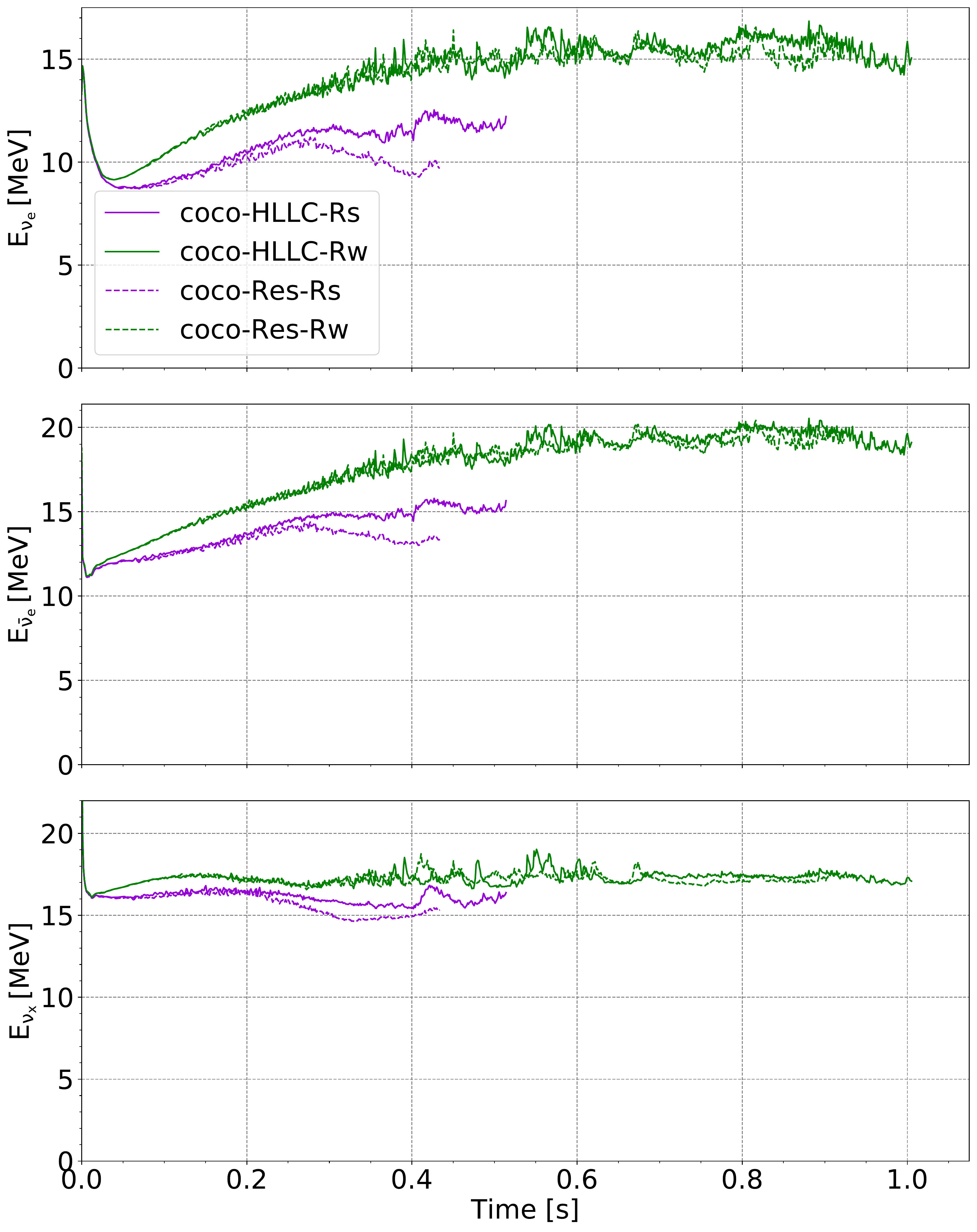}
    \caption{
    Comparison of the mean energies of
    electron neutrinos ($\nu_\mathrm{e}$, top row), electron antineutrinos ($\bar{\nu}_\mathrm{e}$, middle) and  heavy-flavour neutrinos ($\nu_\mathrm{x}$, bottom)
for the standard-resolution coco-HLLC models and the high-resolution coco-Res models,
 measured as spherical averages at a radius of $2000\,\mathrm{km}$
    }
    \label{fig:neutrinoEneRes}
\end{figure}

Unsurprisingly, the variations in PNS surface structure among the strong-field \textsc{CoCoNuT} models are also reflected in the neutrino emission.
Neutrino luminosities and mean energies for all neutrino species for the weak- and strong-field HLLC models at different resolution are shown in 
Figures~\ref{fig:neutrinoLumRes} and
\ref{fig:neutrinoEneRes}, respectively.
 The neutrino luminosities hardly show any resolution dependence, and neither do the mean energies for the weak-field models.
However, the onset of disk inflation in model coco-Res-Rs translates into a marked divergence of the neutrino energies of all flavours from model coco-HLLC-Rs. At the end of the simulation, the mean energies of the electron flavour neutrinos are about
$2\, \mathrm{MeV}$ lower, which is comparable to the differences between \textsc{Alcar} and \textsc{CoCoNuT}, and consistent with the significant expansion of the PNS radius in coco-Res-Rs.

\subsection{Nucleosynthesis Conditions}
Finally, we show histograms of
the $Y_\mathrm{e}$-distribution of the ejecta in model coco-Res-Rs and coco-Res-Rw in Figure~\ref{fig:YeHist}.
As in Section~\ref{nucleosynthesis}, the distribution
is computed for the unbound outflowing material at $\mathrm{\approx} 0.40\,\mathrm{s}$ for the strong-field case and at $\mathord{\approx} 0.85\,\mathrm{s}$ for the weak-field case.

We notice much more extended tails in the high-resolution models.
In model coco-Res-Rs, the neutron-rich tail
reaches down about as far as in the \textsc{Alcar} model to $0.21$, even though the amount of very neutron-rich material is still smaller than in \textsc{Alcar}. In model coco-Res-Rw, the proton-rich tail is now also quite flat as in model Alcar-Rw, only that a numerical cut-off is applied at $Y_\mathrm{e}=0.7$, corresponding to the upper boundary of the low-density NSE table that is used above temperatures of $4.5\, \mathrm{GK}$. However, the ejecta in model coco-Res-Rw still cluster more strongly at a slightly proton-rich value
of $0.7$ than in Alcar-Rw. 

It is noteworthy that the impact of the grid resolution and the Riemann solver on the $Y_\mathrm{e}$-distribution is of a similar magnitude as the differences between the \textsc{CoCoNuT} and \textsc{Alcar} models, i.e., between codes with independent and rather dissimilar neutrino transport schemes. While the impact on  actual yields would have to be evaluated by detailed post-processing of the nucleosynthesis, the sensitivity of the 
$Y_\mathrm{e}$-distribution suggests that r-process yields from current magnetorotational supernova models may still be subject to substantial quantitative uncertainties as these hinge on the neutron-rich tail of the ejecta distribution.

\section{Conclusions} \label{conclusion}
 We conducted the first code comparison of 2D MHD supernova models with the two independent codes \textsc{CoCoNuT-FMT} and \textsc{Aenus-Alcar}.
Starting from the same initial conditions, we simulated the collapse and explosion of a rapidly rotating helium star progenitor model with an initial helium star mass of $35 M_\odot$
\citep{Woosley2006} for two initial magnetic field configurations with the same dipole geometry but different initial field strengths.
The initial poloidal and toroidal field strengths are chosen as
$\mathrm{B_\mathrm{pol,tor} = 10^{12}G}$ in the strong-field case and as $\mathrm{B_\mathrm{pol,tor} = 10^{10}G}$ in the weak-field case. For the \textsc{CoCoNuT-FMT} code, we also investigated the impact of different Riemann solvers (HLLC and HLLE) and the grid resolution \textsc{CoCoNuT-FMT} in a minimal resolution study.

We find overall qualitative agreement between the codes. Both in the strong-field case and in the weak-field case \textsc{CoCoNuT} and \textsc{Alcar} yield explosions at similar times with broadly similar shock trajectories.
Nonetheless, quantitative differences emerge upon closer inspection. Even within the same code, there is an appreciable sensitivity to numerical details like the Riemann solver and the grid resolution, although the various \textsc{CoCoNuT} models are still more similar to each other than to the corresponding \textsc{Alcar} run.

Notable differences include a more rapid explosion in the strong-field \textsc{Alcar} model compared to  \textsc{CoCoNuT-FMT}, which translates into a significantly higher explosion energy. At the end of the simulations, the explosion energies still differ by tens of percent, even though the relative difference may shrink on longer time scales. Similarly, large differences in explosion energy are found for the weak-field models, although these may be impacted by stochastic model variations. Closer inspection reveals differences that can be traced to, or plausibly be associated with the disparate numerical schemes used in the two codes. The treatment of the innermost $10\, \mathrm{km}$ in \textsc{CoCoNuT-FMT} as a quasi-spherical, uniformly rotating region leads to systematic differences in field amplification in the PNS. The two codes also show systematic differences in the PNS surface structure, with \textsc{Alcar} models tending to have a more ``fluffy'' PNS surface or accretion disk around the PNS core. The specific angular momentum profiles and the field amplification in the PNS surface region are also significantly different. The different PNS structure is reflected in the neutrino emission and explains the tendency towards smaller electron-flavour neutrino mean energies in \textsc{Alcar}. However, difference in the neutrino emission also arise due to subtle differences in the infall dynamics and hence the mass accretion rate, and due to the disparate neutrino transport methodology.

We also compared the composition of the ejecta in the various models, specifically the distribution of the electron fraction $Y_\mathrm{e}$ as a crucial factor that decides, e.g., about the production of r-process material in MHD-driven supernovae. The \textsc{CoCoNuT} and \textsc{Alcar} models are qualitatively in agreement in that they produce considerable neutron-rich material in the strong-field case and predominantly proton-rich material in the weak-field case. However, the width of the $Y_\mathrm{e}$-distribution and the amount of very neutron-rich or very proton-rich material varies noticeably
between \textsc{CoCoNuT} and \textsc{Alcar}, and is also quite sensitive to the numerical resolution and the Riemann solver. 

In contrast to 1D comparisons of supernova codes \citep{Liebendorfer2005,Mueller2010,OConnor2018} and multi-D code comparisons for non-rotating, non-magnetised progenitors \citep{Mueller2015,Cabezon2018,Just2018}, disentangling and eliminating the causal factors behind discrepant results proves much more difficult. Therefore no attempt has been made to achieve perfect agreement between the two codes. It is perhaps rather astonishing that despite all the differences, the most coarse-grain metrics of the dynamics like the shock propagation and the explosion energy, and the $Y_\mathrm{e}$-distribution still end up in qualitative agreement.

Our comparison study suggests that MHD supernova models in axisymmetry are quite sensitive to details of the numerical implementation, resolution, and the employed microphysics, and therefore still subject to significant uncertainties.
While they appear robust on a qualitative level, there is clearly not the same level of convergence between codes that has been reached in 1D supernova simulations \citep{Liebendorfer2005,Mueller2010,OConnor2018}. 

Of necessity, our work is only a first step in bracketing and reducing the uncertainties in MHD supernova models. Follow-up studies are highly desirable. Comparison studies with reduced model complexity, e.g., without neutrino transport in the post-bounce phase and an identical treatment of the collapse phase, may be useful to specifically identify and iron out differences between MHD solvers, rather than dealing with a coupled radiation MHD problem. Focusing on MHD only would also make it easier to conduct a comparison in 3D rather than in 2D, which will be critical to study aspects like dynamo field amplification and the kink instability in jet-driven explosions \citep{Moesta_2014,Kuroda2020}. On the other hand, the impact of uncertainties in MHD supernova models with neutrino transport should also be explored in more depth. In particular, it would be desirable to investigate the implications for detailed nucleosynthesis yields rather than just considering the $Y_\mathrm{e}$-distributions in our current study. While MHD supernova simulations still need to further mature in many respects, e.g., by considering more realistic initial magnetic field configurations \citep{Varma2021}, further code comparisons will be an important tool on the long road towards a robust understanding of the role of magnetic fields in the explosions of massive stars.

\section*{Acknowledgements}
We thank A.~Heger, D.~Price and S.~Campbell for useful discussions.
BM acknowledges support by ARC Future Fellowship FT160100035.  
MO acknowledges support from the Spanish Ministry
of Science, Education and Universities (PGC2018-095984-B-I00)
and the Valencian Community (PROMETEU/2019/071), the European Research Council under
grant EUROPIUM-667912, and from the Deutsche Forschungsgemeinschaft
(DFG, German Research Foundation) -- Projektnummer
279384907 -- SFB 1245 as well as from the Spanish Ministry
of Science via the Ramón y Cajal programme (RYC2018-024938-I).
Part of this work has been performed using computer time allocations from Astronomy Australia Limited's ASTAC scheme,  the National Computational Merit Allocation Scheme (NCMAS), and an Australasian Leadership computing grant on the NCI NF supercomputer Gadi. This research was supported by
resources provided by the Pawsey Supercomputing Centre, with funding from the Australian Government and the Government of Western Australia.

\section*{Data Availability}
The data from our simulations will be made available upon reasonable requests made to the authors.


\bibliographystyle{mnras}
\bibliography{library}
\bsp	
\label{lastpage}
\end{document}